%% file: main.tex
\documentclass[preprint,12pt,letterpaper]{elsarticle}

\usepackage{amssymb}
\usepackage{amsthm}
\usepackage{amsmath,mathrsfs}
\usepackage{algorithm}
\usepackage{algorithmicx}
\usepackage{algpseudocode}
\usepackage{color}
\usepackage{multirow}

\biboptions{sort&compress}
\newdefinition{rmk}{Remark}
\journal{Journal of Computational Physics}

\newtheorem{cor}{Corolary}
\newcommand{\ord}[1]{\ensuremath{\mathcal{O}(#1)}}

\newcommand{\trd}{3$^{\textrm{rd}}$}
\newcommand{\fth}{4$^{\textrm{th}}$}
\newcommand{\ie}{\emph{i.e.}}
\newcommand{\eg}{\emph{e.g.}}

\newcommand{\myremarkend}{~\hfill$\clubsuit\/$}

\newcommand{\rrrc}[1]{{\color{red}#1}}
\renewcommand{\rrrc}[1]{#1} 
\newcommand{\reva}[1]{{\color{blue}#1}}
\newcommand{\revb}[1]{{\color{cyan}#1}}
\newcommand{\revc}[1]{{\color{red}#1}}
\newcommand{\revd}[1]{{\color{magenta}#1}}
\renewcommand{\reva}[1]{#1}
\renewcommand{\revb}[1]{#1}
\renewcommand{\revc}[1]{#1}
\renewcommand{\revd}[1]{#1}

\begin{document}

\begin{frontmatter}



\title{High order solution of \rrrc{Poisson problems with
       piecewise constant coefficients and interface jumps}}

\author[mit-aa]{Alexandre Noll Marques}
\author[mcgill]{Jean-Christophe Nave}
\author[mit-math]{Rodolfo Ruben Rosales}

\address[mit-aa]{Department of Aeronautics and Astronautics,
                 Massachusetts Institute of
                 Technology\\Cambridge, MA 02139-4307}
\address[mcgill]{Department of Mathematics and Statistics,
                 McGill University\\Montreal,
                 Quebec H3A 2K6, Canada}
\address[mit-math]{Department of Mathematics,
                   Massachusetts Institute of
                   Technology\\Cambridge, MA 02139-4307}

\input{0abstract}

\begin{keyword}
 Poisson equation \sep
 Correction Function Method \sep
 Interface jump \sep
 High oder \sep
 Immersed Method \sep
 Embedded mesh

\PACS 47.11-j \sep 47.11.Bc

\MSC[2010] 76M20 \sep 35N06



\end{keyword}

\end{frontmatter}


\input{1intro}
\input{2problem}
\input{3potential}
\input{4solution}
\input{5results}
\input{6conclusion}

\input{acknowledgements}




\bibliographystyle{elsarticle-num}
\bibliography{biblioMod}







\end{document}

%% file: 0abstract.tex
\begin{abstract}
We present a fast and accurate algorithm to solve Poisson
problems in complex geometries, using regular Cartesian
grids. We consider a variety of configurations, including
Poisson problems with interfaces across which the solution
is discontinuous (of the type arising in multi-fluid flows).
The algorithm is based on a combination of the Correction
Function Method (CFM) and Boundary Integral Methods (BIM).
Interface and boundary conditions can be treated in a fast
and accurate manner using boundary integral equations, and
the associated BIM. Unfortunately, BIM can be costly when
the solution is needed everywhere in a grid, \eg~fluid flow
problems. We use the CFM to circumvent this issue. The
solution from the BIM is used to rewrite the problem as a
series of Poisson problems in rectangular domains --- which
requires the BIM solution at interfaces/boundaries only.
These Poisson problems involve discontinuities at
interfaces, of the type that the CFM can handle. Hence we
use the CFM to solve them (to high order of accuracy) with
finite differences and a Fast Fourier Transform based fast
Poisson solver. We present 2-D examples of the algorithm
applied to Poisson problems involving complex geometries,
including cases in which the solution is discontinuous.
We show that the algorithm produces solutions that converge
with either \trd\ or \fth\ order of accuracy, depending on
the type of boundary condition and solution discontinuity.
\end{abstract}

%% file: 1intro.tex
\section{Introduction.} \label{sec:intro}
In this paper we present a fast and accurate numerical
algorithm to solve Poisson problems in complex geometries,
using regular Cartesian grids. It can be applied to a wide
variety of Poisson problems, particularly those with
interfaces across which jump conditions are prescribed for
both the solution and its weighted normal derivatives. The
solution to problems of this type is of fundamental
importance in, for example, the description of fluid flows
separated by interfaces (\eg\/~the interfaces between
immiscible fluids, or fluids separated by a membrane). The
algorithm is based on the combination of the Correction
Function Method (CFM)~\cite{marques:11} and boundary
integral formulations of the Laplace
equation~\cite{mikhlin:57, atkinson:97, mclean:00}.

Standard finite differences discretizations cannot be
directly used in the vicinity of a discontinuity interface.
There is a variety of algorithms that have been proposed
for dealing with problems of this type (a brief review of
the literature is included below). In this paper we will
employ the CFM. The CFM produces accurate solutions by
introducing smooth extensions (via the correction function),
across the interface, of the solution on each side. These
extensions are then used to complete the finite differences
stencils straddling the interface. This produces corrections
which affect only the right hand side of the linear system
that would have resulted in the problem without the
interface. The idea is similar to the one used by the
Immersed Boundary Method~\cite{peskin:77}, the Immersed
Interface Method~\cite{leveque:94}, and the Ghost Fluid
Method~\cite{fedkiw:99}. The difference is that in the CFM
the corrections are not computed using expansions. Instead,
the correction function is used, which is defined as the
solution to a PDE, and can be separately solved with any
desired accuracy. As discussed in \cite{marques:11}, the CFM
can be employed to solve the Poisson equation with an
arbitrary immersed interface, across which the solution
obeys appropriate jump conditions, in a regular Cartesian
grid.

However, the CFM~\cite{marques:11} imposes restrictions on
the type of interface jump conditions that it allows, which
are not (generally) satisfied by the problems that arise in
applications.  In this paper we show how to remove these
restrictions. To do so we build on an idea introduced by
Mayo~\cite{mayo:84} to replace the original problem by
sub-problems, each of which can be solved with the CFM. A
brief summary of the process, described in detail
in~\S\ref{sec:solution}, follows.

We write the solution as the sum of two components. The
first component satisfies the Poisson problem in which the
jump in the normal derivative across the interface is set to
vanish. By construction this first problem can be solved
using the CFM. The second component solves the ``deficit''
problem, a Laplace equation which takes care of the jumps in
the normal derivatives at the interface. This second problem
can be solved using a boundary integral method
(BIM)~\cite{mikhlin:57, mayo:84, atkinson:97}. However, we
do not use the BIM to compute the solution at every grid
point, as this would be too costly. Instead we use the potential
densities (at the interface) from the BIM, to write a third
problem which the second component satisfies, which can be
efficiently solved with the CFM.

Furthermore, in the steps where the CFM is used, the solution
domain is embedded into a rectangular box. As a consequence,
the linear algebraic systems that a finite differences
discretization and the CFM produce, can be efficiently
solved using Fast Fourier transform (FFT)
techniques\,\footnote{No
spectral methods involved. The discrete linear equations can
be solved with FFT.}.
Effectively, the problem is reduced to:
 (i) solving two standard (no interfaces, which are removed
     by the CFM formulation) Poisson problems in a box,
     with Dirichlet boundary conditions at the box boundary,
     and
(ii) using a BIM to compute potential densities at the
     interfaces and the boundary.
The overall solution procedure is fast:
the costliest steps involve fast solutions of boundary
integral equations~\cite{rokhlin:85, canning:92, nabors:94},
and an FFT-based fast Poisson solver.

Finally, in principle the BIM and the CFM can be implemented
to any order of accuracy. Therefore, the present algorithm
can be made as accurate as one desires.

There exists a vast literature on immersed methods for the
Poisson equation and related problems.
By immersed methods here we mean: methods where the entire
solution domain (normally involving complex geometries) is
immersed into a regular Cartesian grid or triangulation.
In particular, the combination of boundary integral
equations and immersed methods was introduced in the seminal
work by Mayo and co-authors~\cite{mayo:84, mayo:jcp:1992,
mckenney:95}.
Recent work of particular interest focuses on Kernel-Free
Boundary Integral Methods~\cite{ying:13}.
This new class may offer better efficiency for 3-D problems.
Some of the other well established methods include the work
by Johansen and Collela~\cite{johansen:98},
\revb{the Immersed Boundary Method~\cite{peskin:77,
roma:1999, lai:00},
the Immersed Interface Method~\cite{leveque:94, leveque:97,
zhang:1997, hou:1997, li:01, lee:03}},
the Ghost Fluid Method~\cite{fedkiw:99, fedkiwetal:99,
liu:00, kang:00, nguyen:01, gibou:07}, and
the Immersed Boundary Smooth Extension
Method~\cite{stein:16}.
\revd{A recent development is the Voronoi Interface
Method~\cite{guittet:2015}, an extension of the Ghost Fluid
Method that achieves second accuracy in the solution and
first order accuracy in the gradient for all regimes.}
The finite element community has also made significant
progress in developing new immersed methods.
For example, the work by Dolbow and Harari~\cite{dolbow:09},
the Extended Finite Element Method~\cite{moes:99},
the Virtual Node Method~\cite{bedrossian:10},
the finite element versions of the Immersed Interface
Method~\cite{li:03, hou:05, li:06, gong:08}, and
the Exact Subgrid Interface Correction
Method~\cite{huh:08}.

\revb{High order (specifically fourth order) implementations of some the methods mentioned above are known.
For example: Ghost Fluid Method~\cite{gibou:2005}, and
combination of BIM and immersed methods~\cite{mayo:2003, mayo:2007}.}

The present paper is inspired by the work by Mayo and
co-authors~\cite{mayo:84, mayo:jcp:1992, mckenney:95}.
We extend some of their general ideas by incorporating the
CFM, which allows us to solve Poisson problems involving
interface jumps, and offers a general framework to develop
high order schemes.

The remainder of the paper is organized as follows.
In~\S\ref{sec:problem} we introduce the general Poisson
problem that we seek to solve. In \S\ref{sec:potential} we
review well established results pertaining to the potential
formulation of the Laplace equation. In \S\ref{sec:solution}
we present the proposed algorithm, and discuss details such
as accuracy and computational cost. In \S\ref{sec:results}
we show examples of the algorithm applied to problems
involving complex geometries and interfaces of
discontinuity, including convergence results. Finally,
\S\ref{sec:conclusion} contains the conclusions.


%% file: 2problem.tex
\section{Definition of the problem} \label{sec:problem}
\subsection{Notation} \label{sec:problem:Notn}
\begin{figure}[htb!]
 \begin{center}
  \includegraphics[width=2.6in]{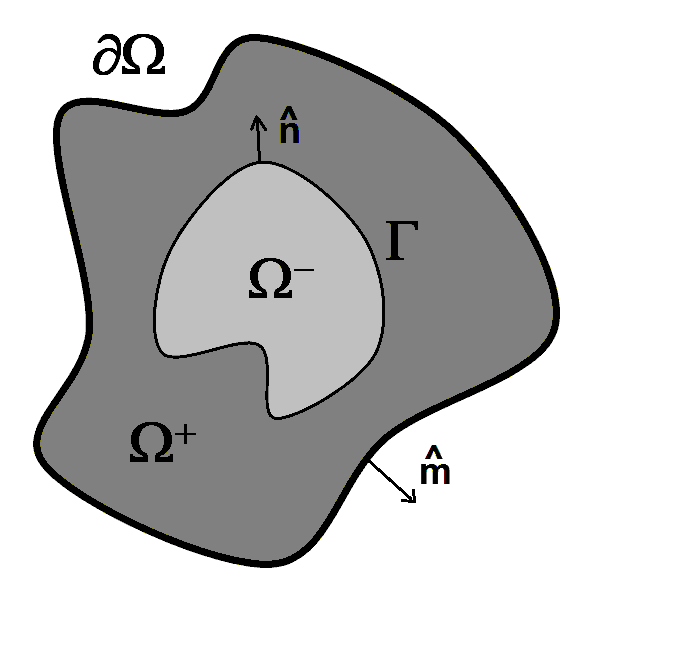}
  \includegraphics[width=2.6in]{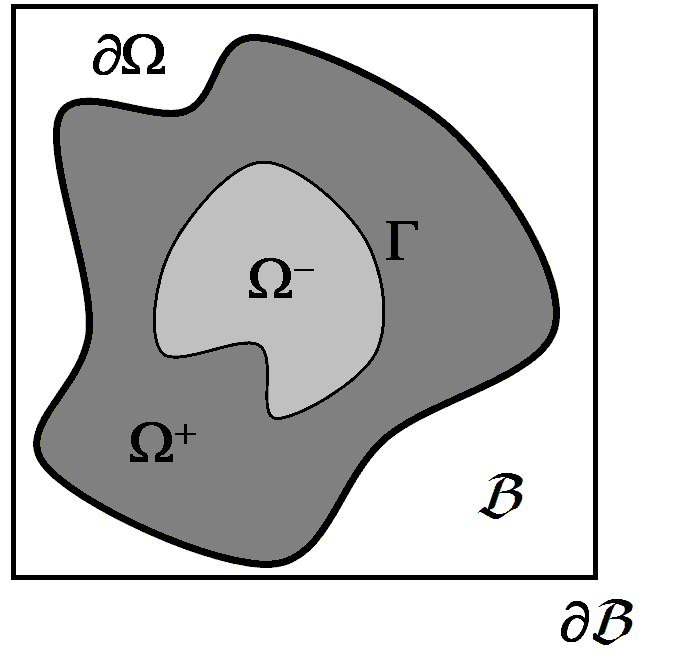}
 \end{center}
 \caption{Left: typical solution domain $\Omega\/$, split by
          an internal interface $\Gamma\/$ across which the
          solution and the equation coefficients jump. Note
          that there is no problem with allowing
          $\Omega^-\/$ to have more than one component.
          Right: the box $\mathcal{B}$ enclosing $\Omega\/$.
          The role of $\mathcal{B}$ is explained
          in~\S\ref{sec:solution}.}
 \label{fig:problem}
\end{figure}

Throughout this paper, $\vec{x} = (x_1\/,\,x_2\/,\,\dots)
\in \mathbb{R}^\nu\/$ is the spatial vector (where
$\nu = 2\/$ or $\nu = 3\/$), $\Delta\/$ is the Laplace
operator defined by
\begin{equation}
 \Delta\/ = \sum_{i=1}^{\nu}
 \dfrac{\partial^2}{\partial x_i^2}\/,
\end{equation}
and $\Omega\/$ is an arbitrary, bounded and open, simply
connected domain in $\mathbb{R}^{\nu}\/$, with a piece-wise
smooth boundary $\partial \Omega\/$. This domain is split
into two sub-domains by a co-dimension 1 surface $\Gamma\/$
disjoint from the boundary $\partial \Omega\/$. We denote
the sub-domain interior to $\Gamma\/$ by $\Omega^-\/$, and
the sub-domain exterior to $\Gamma\/$ by $\Omega^+\/$. The
situation we have in mind is best described by a picture:
see figure~\ref{fig:problem}.

Let $\hat{n}\/$ denote the unit vector normal to $\Gamma\/$,
pointing towards $\Omega^+\/$, and let $u\/$ be a function
defined in a neighborhood of $\Gamma\/$. Then
\begin{align}
 u_n &= \hat{n}\cdot\vec{\nabla}u =
 \hat{n}\cdot(u_{x_1}\/,\,u_{x_2}\/,\,\dots)
 & \mbox{for}\;\vec{x} &\in \Gamma
\end{align}
denotes the derivative of $u\/$ in the direction of
$\hat{n}\/$. Similarly, $\hat{m}\/$ and $u_m\/$ denote the
outer normal vector to the boundary $\partial \Omega\/$,
and the corresponding directional derivative.

Furthermore, we are interested in problems that involve
jumps. Let \linebreak $\vec{x} = \vec{x}_0 +
\epsilon\hat{n}(\vec{x}_0)\/$, where $\vec{x}_0 \in
\Gamma\/$. Then, we denote jumps across the surface
$\Gamma\/$ at $\vec{x}_0\/$ by
\begin{equation}\label{eq:jump}
 [u(\vec{x}_0)]_{\Gamma} =
 \lim_{\epsilon \uparrow 0} u(\vec{x})
 - \lim_{\epsilon \downarrow 0} u(\vec{x})\/,
\end{equation}
and the corresponding mean value by
\begin{equation}\label{eq:mean}
 \langle u(\vec{x}_0) \rangle_{\Gamma} =
 \dfrac{1}{2} \lim_{\epsilon \uparrow 0} u(\vec{x})
 + \dfrac{1}{2} \lim_{\epsilon \downarrow 0} u(\vec{x})\/.
\end{equation}
Similar expressions apply to jumps across
$\partial\Omega\/$.

\subsection{Poisson problems with an interface}
\label{sec:problem:Eqns}
Our objective is to solve Poisson problems of the form:
\begin{subequations}\label{eq:poisson}
 \begin{align} 
   \revc{\vec{\nabla}\cdot(\beta\/^+ \vec{\nabla} u(\vec{x}))} & =
   f^+(\vec{x}) &\mathrm{for}\;\;
   \vec{x} &\in \Omega\/^+,
   \label{eq:poisson:eqp}\\ \rule{0mm}{1.2em}
   \revc{\vec{\nabla}\cdot(\beta\/^- \vec{\nabla} u(\vec{x}))} & =
   f^-(\vec{x}) &\mathrm{for}\;\;
   \vec{x} &\in \Omega\/^-,
   \label{eq:poisson:eqm}
 \end{align}
\end{subequations}
with the jump conditions
\begin{subequations}\label{eq:jumpcond}
 \begin{align}
   [u(\vec{x})]_{\Gamma} & = a(\vec{x})
   &\mathrm{for}\;\; \vec{x} &\in \Gamma\/,
   \label{eq:jumpcond:du}\\ \rule{0mm}{1.2em}
   \bigl[\beta\/u_n(\vec{x})\bigr]_{\Gamma} & = b(\vec{x})
   &\mathrm{for}\;\; \vec{x} &\in \Gamma\/,
   \label{eq:jumpcond:dun_1}
 \end{align}
\end{subequations}
and either Dirichlet or Neumann boundary conditions:
\begin{subequations}\label{eq:bc}
 \begin{align}
  u(\vec{x}) & = g_D(\vec{x})
  &\mathrm{for}\;\; \vec{x} &\in \partial\Omega\/,
  \label{eq:bc:d}\\ \rule{0mm}{1.2em}
  \mathrm{or} \qquad u_m(\vec{x}) & = g_N(\vec{x})
  &\mathrm{for}\;\; \vec{x} &\in \partial\Omega\/.
  \label{eq:bc:n}
 \end{align}
\end{subequations}
Here $\beta^{\pm}\/$ are positive constants, $f^{\pm} \colon
\Omega^{\pm} \mapsto \mathbb{R}\/$, $a\/$, $b \colon \Gamma
\mapsto \mathbb{R}$, and \linebreak
$g_D\/$, $g_N \colon \partial \Omega \mapsto \mathbb{R}$ are
some given functions.\,\footnote{As
 discussed in \cite{marques:11}, how smooth these functions
 need to be is tied up to how accurate an approximation is
 desired. For instance, for a \fth\ order algorithm, we must
 have \linebreak
 $f^{\pm} \in C^2(\Omega^{\pm})\/$, $a \in C^4(\Gamma)\/$,
 $b \in C^3(\Gamma)\/$, $g_D \in C^4(\partial \Omega)\/$,
 and $g_N \in C^3(\partial \Omega)\/$.}
Furthermore, in the case of Neumann boundary condition,
equation~\eqref{eq:bc:n}, we further assume that the
following compatibility condition is satisfied
\begin{equation}\label{eq:compatibility}
 \int_{\Omega}f\,d\/V =
 \int_{\partial\Omega} \beta^+ g_N\,d\/S 
 - \int_{\Gamma} b\,d\/S\/.
\end{equation}

In the development of the algorithm, it is convenient to
rewrite \eqref{eq:jumpcond:dun_1} in terms of
$[u_n]_{\Gamma}\/$. We use the following identity
\begin{equation}\label{eq:pq}
 [pq] = \langle p \rangle[q] + [p]\langle q \rangle,
\end{equation}
to write the alternative form
\begin{align}\label{eq:jumpcond:dun}
 [u_n]_{\Gamma} + \lambda \langle u_n \rangle_{\Gamma} & =
 b(\vec{x})/\langle\/\beta\/\rangle_{\Gamma}\/.
\end{align}
where
$\lambda = [\beta]_{\Gamma}/\langle\beta\rangle_{\Gamma}\/$.

The problems above are known as \emph{interior} Poisson
problems~\cite{atkinson:97} because we solve for $u\/$ in
the interior of the boundary $\partial \Omega\/$. In
contrast, in \emph{exterior} Poisson problems we solve for
$u\/$ in the exterior of $\partial \Omega\/$.
Exterior problems can be solved with techniques analogous to
the ones used in this paper, but we will not consider them
in here.

We also discuss the solution to \emph{open space} problems,
where the solution is defined everywhere, \ie\ $\Omega =
\mathbb{R}^{\nu}\/$. In this case, we assume that $f\/$ is
compact support and denote the support of $f\/$ by
$\Omega_f\/$.
\revc{Furthermore, to guarantee uniqueness, we also assume
that the solution exhibits the following asymptotic behavior
\begin{subequations}\label{eq:infty}
 \begin{align}
  u(\vec{x})
  &\; \sim \; \dfrac{F}{2\pi}\ln\lvert\vec{x}\rvert,
  &\mbox{for $\lvert\vec{x}\rvert \gg 1$ and $\nu=2$,}\\
  u(\vec{x})
  &\; \sim \;
   \dfrac{F}{(\nu-2)A_{\nu}\lvert\vec{x}\rvert^{\nu-2}},
  &\mbox{for $\lvert\vec{x}\rvert \gg 1$ and $\nu=3$,}
 \end{align}
\end{subequations}
where $A_{\nu}\/$ is  area of the unit sphere in
$\mathbb{R}^\nu\/$, and}
\begin{equation}\label{eq:F}
 F = \dfrac{1}{\beta^+}\int_{\Omega^+} f^+\,d\/V
   + \dfrac{1}{\beta^-}\int_{\Omega^-} f^-\,d\/V
   + \int_{\Gamma} [u_n]_{\Gamma}dS\/.
\end{equation}

\begin{rmk} \label{rmk:problem:ccoeff}
 In the literature the case when $\beta^+ = \beta^-\/$ is
 often called the \emph{constant coefficients} case, while
 $\beta^+ \neq \beta^-\/$ is referred to as the
 \emph{discontinuous coefficients} case. There is also the
 case where $\beta^{\pm} = \beta^{\pm}(\vec{x})\/$, known
 as \emph{variable coefficients} case. This last case is
 not treated in this paper.\myremarkend
\end{rmk}

\begin{rmk} \label{rmk:problem:neumann}
 The solution to (\ref{eq:poisson}--\ref{eq:jumpcond}) with
 Neumann boundary condition~\eqref{eq:bc:n} (interior
 Neumann problem) is defined up to an arbitrary additive
 constant. There is a number of techniques that can be used
 to obtain a solution to this problem~\cite{mayo:84,
 atkinson:97, romate:93}.  In~\S\ref{sub:solution:singular}
 we discuss how we use the Generalized Minimum Residual
 Method (GMRES)~\cite{trefthen:97} to obtain accurate
 solutions to the singular systems of equations that arise
 in this context.\myremarkend
\end{rmk} 

\subsection{The large $\beta\/$ ratio limit}
\label{sub:problem:poorcond}
Many applications in multiphase flows involve the solution
of problems like the ones above, with a large ratio between
the coefficients $\beta^\pm\/$. For example, in the
simulation of multi-phase flows, $\beta\/$ is the reciprocal
of the fluid density, and hence for air-water interfaces the
ratio of $\beta^\pm\/$ is $10^3\/$. As we will demonstrate,
the solution algorithm presented in~\S\ref{sec:solution} is
general enough to deal with such ratios, and larger.

However, the situation where $\beta^-/\beta^+ \gg 1\/$
involves a subtlety. In the limit $\beta^-/\beta^+ \to
\infty\/$, equation \eqref{eq:jumpcond:dun} yields
$u^-_n = 0\/$, where $u^-\/$ denotes the restriction of
$u\/$ to $\Omega^-\/$. Thus $u^-\/$ becomes the solution to
a Poisson problem, with a Neumann boundary condition on
$\Gamma = \partial \Omega^-\/$. Hence $u^-\/$ is defined
only up to an arbitrary additive constant. For this reason,
when $\beta^-/\beta^+\/$ is large, the Poisson problem in
(\ref{eq:poisson}--\ref{eq:jumpcond}) becomes poorly
conditioned.
\emph{This issue is intrinsic to the problem being
solved, independent of the numerical algorithm used.}

In \S\ref{sub:solution:poorcond} we discuss an approach to
to remove the poor conditioning in the context of the
algorithm described here.


%% file: 3potential.tex
\section{Potential formulation}\label{sec:potential}
In this section we review a few basic results pertaining to
the fundamental solution of the Laplace equation, as they
will be important to design our method. A detailed
discussion of these results are found in many textbooks on
the subject, \eg\ \cite{atkinson:97, evans:98, mclean:00}.

Let $\Phi = \Phi(\vec{x}\/,\,\vec{y})\/$ denote the 
\emph{fundamental solution of the Laplace equation}, defined
by
\begin{equation} \label{eq:fundamental}
 \Delta\,\Phi = \delta(\vec{x}-\vec{y})\/,
\end{equation}
where $\delta\/$ denotes the Dirac delta distribution. In
particular\vspace*{-0.4em}
\begin{itemize}
 \item
  \parbox{2.70in}{in 2-D,}
  ${\displaystyle
    \Phi = \dfrac{1}{2\,\pi}\,
    \ln(\lvert\vec{x}-\vec{y}\rvert)}\/$,
 \item \vspace*{-0.4em}
  \parbox{2.70in}{in 3-D,}
  ${\displaystyle
    \Phi = -\dfrac{1}
    {4\,\pi\,\lvert\vec{x}-\vec{y}\rvert}}\/$,
 \item \vspace*{-0.4em}
  \parbox{2.70in}{generally, in  $\nu \ge 3\/$
   dimensions,}
   ${\displaystyle
     \Phi = -\dfrac{1}
     {(\nu-2)\,A_{\nu}\,
      \lvert\vec{x}-\vec{y}\rvert^{\nu-2}}}\/$,
   \\where $A_{\nu} =\/$ area of the unit sphere in
   $\mathbb{R}^\nu\/$.
\end{itemize}
For $\vec{x} \in \Gamma\/$ (or $\vec{x} \in \partial
\Omega\/$), $\Phi_n\/$ (or $\Phi_m\/$) denotes the normal
derivative of $\Phi(\vec{x}\/,\,\vec{y})\/$ with respect to
$\vec{x}\/$. Furthermore, for $\vec{x} \in \Gamma\/$ and
$\vec{y} \in \partial \Omega\/$, $\Phi_{n\/m}\/$ is the
object:
\begin{equation} \label{eq:fundamentalnm}
 \Phi_{n\/m} =
 \sum_{i\,j} n_i\,m_j\,
 \dfrac{\partial^2}{\partial x_i\,\partial y_j}\,\Phi\/.
\end{equation}
Fairly straightforward, but tedious, calculations show that
the singularity in $\Phi\/$ is such that 
\begin{equation} \label{eq:phiint}
 \text{For $\vec{x} \in \Gamma\/$, $\Phi\/$ and $\Phi_n\/$
       are integrable as functions of $\vec{y}\/$ in
       $\Gamma\/$.}
\end{equation}

Many common boundary integral formulations of the Laplace
equation follow from Green's third
identity~\cite{mclean:00}. This identity shows that the
solution to the Laplace equation can be written as a
combination of single and double layer potentials. For ease
of presentation, here we describe single and double layer
potentials based on $\Gamma\/$, but similar results apply to
$\partial \Omega\/$.

A single layer potential solution based on $\Gamma\/$ has
the form
\begin{equation} \label{eq:sl}
 u^{\text{SL}}(\vec{x}) =
 \int_{\Gamma} \rho(\vec{y})
 \Phi(\vec{x}\/,\,\vec{y})\,d\/S\/,
\end{equation}
where the integration is taken over $\vec{y} \in \Gamma\/$,
and $\rho\/$ is the single layer density defined over
$\Gamma\/$. A single layer potential is a solution of the
Laplace equation everywhere in $\mathbb{R}^{\nu}\/$, and it
satisfies the following jump conditions
\begin{subequations}\label{eq:sl:jumps}
 \begin{align}
  [u^{\text{SL}}(\vec{x})]_{\Gamma} &= 0\/,
  \label{eq:sl:jumps:u}\\ \rule{0mm}{1.2em}
  [u^{\text{SL}}_n(\vec{x})]_{\Gamma}
  &= \rho(\vec{x})\/.
  \label{eq:sl:jumps:un}
 \end{align}
\end{subequations}
\rrrc{Furthermore,} 
let $\vec{x} = \vec{x}_0 + \epsilon\,\hat{n}(\vec{x}_0)\/$,
where $\vec{x}_0 \in \Gamma\/$. Then
\begin{equation} \label{eq:sl:lim:u}
 \lim_{\epsilon \uparrow 0}
 u^{\text{SL}}(\vec{x}) =
 \lim_{\epsilon \downarrow 0}
 u^{\text{SL}}(\vec{x}) =
 \int_{\Gamma} \rho(\vec{y})
 \Phi(\vec{x}_0\/,\,\vec{y})\,d\/S\/,
\end{equation}
and
\begin{subequations} \label{eq:sl:lim:un}
 \begin{align}
  \lim_{\epsilon \uparrow 0}
  u^{\text{SL}}_n(\vec{x}) & =
  \phantom{-} \dfrac{1}{2}\rho(\vec{x}_0) +
  \int_{\Gamma}
  \rho(\vec{y})\Phi_n(\vec{x}_0\/,\,\vec{y})\,d\/S\/,
  \label{eq:sl:lim:unm}\\ \rule{0mm}{1.9em}
  \lim_{\epsilon \downarrow 0}
  u^{\text{SL}}_n(\vec{x}) & =
  - \dfrac{1}{2}\rho(\vec{x}_0) +
  \int_{\Gamma}
  \rho(\vec{y})\Phi_n(\vec{x}_0\/,\,\vec{y})\,d\/S\/.
  \label{eq:sl:lim:unp}
 \end{align}
\end{subequations}
In open space problems we are also interested in the
behavior of the single layer potential at infinity.
\rrrc{It is straightforward to show that}
\begin{equation} \label{eq:sl:infty}
 \lim_{\lvert \vec{x} \rvert \to \infty}
 u^{\text{SL}}(\vec{x}) =
 \left\{\begin{array}{ll}
  \dfrac{1}{2\pi}\ln\lvert\vec{x}\rvert
  \displaystyle
  \int_{\Gamma} \rho(\vec{y})\,d\/S\/, &
  \mbox{in $\nu=2$ dimensions,}\\ \rule{0mm}{2em}
  -\dfrac{1}
  {(\nu-2)\mathcal{A}_{\nu}\lvert\vec{x}\rvert^{\nu-2}}
  \displaystyle
  \int_{\Gamma} \rho(\vec{y})\,d\/S\/, &
  \mbox{in $\nu \ge 3$ dimensions.}
 \end{array}\right.
\end{equation}

A double layer potential solution based on $\Gamma\/$ has
the form
\begin{equation} \label{eq:dl}
 u^{\text{DL}}(\vec{x}) =
 \int_{\Gamma} \mu(\vec{y})
 \Phi_n(\vec{x}\/,\,\vec{y})\,d\/S\/,
\end{equation}
where $\mu\/$ is the double layer density defined over
$\Gamma\/$. A double layer potential solution satisfies the
following jump conditions
\begin{subequations}\label{eq:dl:jumps}
 \begin{align}
  [u^{\text{DL}}(\vec{x})]_{\Gamma} &= -\mu(\vec{x})\/,
  \label{eq:dl:jumps:u}\\ \rule{0mm}{1.2em}
  [u^{\text{DL}}_n(\vec{x})]_{\Gamma} &= 0\/.
  \label{eq:dl:jumps:un}
 \end{align}
\end{subequations}
In addition, \rrrc{with
$\vec{x}=\vec{x}_0+\epsilon\,\hat{n}(\vec{x}_0)\/$,}
\begin{subequations} \label{eq:dl:lim:u}
 \begin{align}
  \lim_{\epsilon \uparrow 0}
  u^{\text{DL}}(\vec{x}) & =
  - \dfrac{1}{2}\mu(\vec{x}_0) +
  \int_{\Gamma} \mu(\vec{y})
  \Phi_n(\vec{x}_0\/,\,\vec{y}) dS\/,
  \label{eq:dl:lim:unm}\\ \rule{0mm}{1.8em}
  \lim_{\epsilon \downarrow 0}
  u^{\text{DL}}(\vec{x}) & =
  \phantom{-} \dfrac{1}{2}\mu(\vec{x}_0) +
  \int_{\Gamma} \mu(\vec{y})
  \Phi_n(\vec{x}_0\/,\,\vec{y}) dS\/.
  \label{eq:dl:lim:unp}
 \end{align}
\end{subequations}
The limits of the normal derivatives of \eqref{eq:dl}
involve generalized functions on $\Gamma\/$. However, these
expressions are not of interest for the algorithm
in~\S\ref{sec:solution}.


%% file: 4solution.tex
\section{Solution method} \label{sec:solution}
In this section we present an algorithm to solve the Poisson
problems introduced in~\S\ref{sec:problem:Eqns}. This
algorithm is based on a combination of the Correction
Function Method (CFM)~\cite{marques:11} and an adaptation of
the boundary integral formulations of the Laplace
equation~\cite{mayo:84,mayo:jcp:1992}. 

The solution is split into two components: $u = v + w\/$.
The first component, $v\/$, is the solution to a Poisson
problem that can be solved with the CFM --- see
remark~\ref{rmk:problem:CFM} below. We present a formal
definition of $v\/$ is in~\S\ref{sub:solution:v}. The second
component, $w\/$, is the solution to the ``deficit''
problem: a Laplace equation that includes the the boundary
and jump conditions that the CFM cannot handle in a direct
fashion. In~\S\ref{sub:solution:w1}--\S\ref{sub:solution:w2}
we show how to solve for $w\/$ using a combination of
boundary integral equations and the CFM.

We introduce the algorithm for interior problems
in~\S\ref{sub:solution:v}--\S\ref{sub:solution:w2}, and
extend the algorithm to open space problems
in~\S\ref{sub:solution:open}.
In~\S\ref{sub:solution:implementation} we comment on a few
details of practical relevance for the implementation of the
algorithm.  In~\S\ref{sub:solution:singular} we address the
issue of how to deal with the singular problem that arises
when Neumann conditions are imposed.
In~\S\ref{sub:solution:poorcond} show how to manage the
poorly conditioned problems that arise when $\beta^- \gg
\beta^+\/$. Finally, in~\S\ref{sub:solution:accuracy} we
discuss the accuracy and efficiency of the algorithm.

\begin{rmk} \label{rmk:problem:CFM}
 The Correction Function Method (CFM)~\cite{marques:11} was
 developed to deal with interface conditions of the form in
 \eqref{eq:jumpcond}, for the case when
 $\beta^- = \beta^+\/$. The CFM offers a framework to solve
 problems of this type, to high order of accuracy, using
 finite differences on a regular Cartesian grid.

 The CFM is based on the computation of a \emph{correction
 function}, which provides smooth and accurate extensions of
 $u^\pm\/$ across $\Gamma\/$. The correction function is
 defined as the solution to a PDE problem and, in principle,
 can be computed to an arbitrary order of accuracy. This
 correction function can then be used to complete finite
 differences discretizations of the Laplace operator, without
 loss of accuracy, for stencils that straddle $\Gamma\/$.
 The CFM produces a discretized linear system whose
 coefficient matrix is the same as the one that arises in
 the absence of interfaces --- the systems differ only by
 their right hand sides. Hence the same linear solvers that
 work for ``standard'' Poisson problems can be used. A \fth\
 order implementation of the CFM can be found
 in~\cite{marques:11}. We use this implementation to obtain
 the results presented in~\S\ref{sec:results}.\myremarkend
\end{rmk} 

\input{4solution41v}
\input{4solution42w1}
\input{4solution43w2}
\input{4solution44open}
\input{4solution45implementation}
\input{4solution46singular}
\input{4solution47poorcond}
\input{4solution48accuracy}


%% file: 4solution41v.tex
\subsection{First solution component}
\label{sub:solution:v}
The first solution component, $v\/$, incorporates the terms
that can be directly solved with the CFM: the
non-homogeneous source term, $f\/$, and the jump
$[u]_{\Gamma} = a\/$. Furthermore, for the sake of
simplicity and efficiency, $v\/$ is defined in a rectangular
box $\mathcal{B}$ that includes $\Omega\/$ --- see
figure~\ref{fig:problem}. Then, $v\/$ is the solution to the
following Poisson problem:
\begin{subequations}\label{eq:v}
 \begin{align}
   \Delta\/v(\vec{x}) &=
   \begin{cases}
    0                      \\ \rule{0mm}{1.1em}
    f^+(\vec{x})/\beta\/^+ \\ \rule{0mm}{1.1em}
    f^-(\vec{x})/\beta\/^-
   \end{cases} &
   \begin{matrix}
    \mathrm{for}\;\;\vec{x} \\ \rule{0mm}{1.1em}
    \mathrm{for}\;\;\vec{x} \\ \rule{0mm}{1.1em}
    \mathrm{for}\;\;\vec{x}
   \end{matrix} &
   \begin{matrix}
    \in \mathcal{B}/\Omega\/,\\ \rule{0mm}{1.1em}
    \in \Omega^+\/,          \\ \rule{0mm}{1.1em}
    \in \Omega^-\/,
   \end{matrix}
   \label{eq:v:eq}\\  \rule{0mm}{1.2em}
   [v]_{\Gamma} &=
   a(\vec{x}) &
   \mathrm{for}\;\;\vec{x} & \in \Gamma\/,
   \label{eq:v:dug}\\ \rule{0mm}{1.1em}
   [v_n]_{\Gamma} &= 0 &
   \mathrm{for}\;\;\vec{x} & \in \Gamma\/,
   \label{eq:v:dung}\\ \rule{0mm}{1.1em}
   [v]_{\partial \Omega} &= 0 &
   \mathrm{for}\;\;\vec{x} & \in \partial \Omega\/,
   \label{eq:v:dupo}\\ \rule{0mm}{1.1em}
   [v_m]_{\partial \Omega} &= 0 &
   \mathrm{for}\;\;\vec{x} & \in \partial \Omega\/,
   \label{eq:v:dumpo}\\ \rule{0mm}{1.1em}
   v(\vec{x}) &= 0
   &\mathrm{for}\;\; \vec{x} &\in \partial\mathcal{B}\/.
   \label{eq:v:bcd}
 \end{align}
\end{subequations}
By construction, \eqref{eq:v} is a problem that can be
solved using the CFM. Furthermore, because the CFM does not
affect the discretization of the Poisson equation, and
\eqref{eq:v} is defined in a rectangular box, $v\/$ can be
computed with a fast Poisson solver based on the FFT.


%% file: 4solution42w1.tex
\subsection{Second solution component: boundary integral
            formulation}\label{sub:solution:w1}
The second solution component, $w\/$, is the solution to the 
``deficit'' problem that follows from subtracting
\eqref{eq:v} from (\ref{eq:poisson}--\ref{eq:bc}):
\begin{subequations}\label{eq:w1}
 \begin{align}
  \Delta\/w(\vec{x}) &= 0
  &\mathrm{for}\;\; \vec{x} &\in \Omega\/,
  \label{eq:w1:eq}\\ \rule{0mm}{1.1em}
  [w]_{\Gamma} &= 0
  &\mathrm{for}\;\; \vec{x} &\in \Gamma\/,
  \label{eq:w1:dwg}\\ \rule{0mm}{1.1em}
  [w_n]_{\Gamma} + \lambda\/\langle w_n \rangle_{\Gamma} &=
  b(\vec{x})/\langle \beta \rangle_{\Gamma} -
  \lambda\/v_n(\vec{x})
  &\mathrm{for}\;\; \vec{x} &\in \Gamma\/,
  \label{eq:w1:dwng}
 \end{align}
\end{subequations}
with boundary conditions
\begin{subequations}\label{eq:w1:bc}
 \begin{align}  
  w(\vec{x}) & = g_D(\vec{x}) - v(\vec{x})
  &\mathrm{for}\;\; \vec{x} &\in \partial\Omega\/,
  \label{eq:w1:bc:d}\\ \rule{0mm}{1.1em}
  \mathrm{or} \qquad
  w_m(\vec{x}) & = g_N(\vec{x}) - v_m(\vec{x})
  &\mathrm{for}\;\; \vec{x} &\in \partial\Omega\/.
  \label{eq:w1:bc:n}
 \end{align}
\end{subequations}

Note that, because equation~\eqref{eq:v} includes the
forcing term $f\/$, the ``deficit'' problem \eqref{eq:w1}
becomes a Laplace problem. An efficient approach to solve a
Laplace equation such as \eqref{eq:w1} is to use a boundary
integral formulation. As discussed in \cite{mclean:00},
Green's third identity guarantees the solution to
\eqref{eq:w1} can be written as
\begin{align}\label{eq:wbif}
 \begin{split}
  w(\vec{x}) &=
  \int_{\Gamma}
  \rho_{\Gamma}(\vec{y})\,\Phi(\vec{x},\,\vec{y})\,d\/S \\
  \rule{0mm}{1.7em}
  &+\int_{\partial \Omega}
  \bigl(\rho_{\partial \Omega}(\vec{y})\,
  \Phi(\vec{x},\,\vec{y}) +
  \mu_{\partial \Omega}(\vec{y})\,
  \Phi_m(\vec{x},\,\vec{y})\bigr)\,d\/S
 \end{split}
 &\mathrm{for}\;\; \vec{x} &\in \Omega\/,
\end{align}
where the integration is taken over $\vec{y}\/$ and $\Phi\/$
is the fundamental solution of the Laplace equation.
Expression~\eqref{eq:wbif} satisfies the Laplace equation
identically. Then, the unknown potential densities ($\rho\/$
and $\mu\/$) result from the jump and boundary conditions.
It follows directly from the formulas
in~\S\ref{sec:potential} that these conditions lead to the
following expressions:

\goodbreak

\bigskip\noindent
\hspace*{0em}\hfill \emph{Equations for the case of
Dirichlet boundary conditions} \hfill\hspace*{0em}
\begin{subequations}\label{eq:wbie:id}
 \begin{align}
  \rho_{\partial \Omega}&(\vec{x}) = 0 \rule{0mm}{1.3em}
  &\mathrm{for}\;\; \vec{x} &\in \partial \Omega\/,
  \rule[-0.9em]{0mm}{1.2em}
  \label{eq:wbie:id:dwnpo}\\  
  \begin{split}
   \rho_{\Gamma}(\vec{x}) &+
   \lambda\,\int_{\Gamma}
   \rho_{\Gamma}(\vec{y})\,
   \Phi_n(\vec{x},\,\vec{y})\,d\/S\,+\\ \rule{0mm}{1.7em}
   & \lambda\,\int_{\partial \Omega}
   \mu_{\partial \Omega}(\vec{y})\,
   \Phi_{nm}(\vec{x},\,\vec{y})\,d\/S
   = \dfrac{b(\vec{x})}{\langle \beta \rangle_{\Gamma}} -
   \lambda\,v_n(\vec{x}) \rule[-1.4em]{0mm}{1.2em}
  \end{split}
  &\mathrm{for}\;\; \vec{x} &\in \Gamma\/,
  \label{eq:wbie:id:dwng}\\
  \begin{split}
   \mu_{\partial \Omega}(\vec{x}) &+
   2\,\int_{\Gamma}
   \rho_{\Gamma}(\vec{y})\,
   \Phi(\vec{x},\,\vec{y})\,d\/S\,+\\\rule{0mm}{1.7em}
   & 2\,\int_{\partial\Omega}
   \mu_{\partial \Omega}(\vec{y})\,
   \Phi_m(\vec{x},\,\vec{y})\,d\/S
   = 2\,(g_D(\vec{x}) - v(\vec{x}))
  \end{split}
  & \mathrm{for}\;\; \vec{x} &\in \partial\Omega\/.
  \label{eq:wbie:id:bc}
 \end{align}
\end{subequations}

\bigskip\noindent
\hspace*{0em}\hfill \emph{Equations for the case of
Neumann boundary conditions} \hfill\hspace*{0em}
\begin{subequations}\label{eq:wbie:in}
 \begin{align}
  \mu_{\partial \Omega}(\vec{x}) &= 0 \rule{0mm}{1.3em}
  &\mathrm{for}\;\; \vec{x} &\in \partial \Omega\/,
  \rule[-0.9em]{0mm}{1.2em}
  \label{eq:wbie:in:dwpo}\\  
  \begin{split}
   \rho_{\Gamma}(\vec{x}) &+
   \lambda\,\int_{\Gamma}
   \rho_{\Gamma}(\vec{y})\,
   \Phi_n(\vec{x},\,\vec{y})\,d\/S\,+\\ \rule{0mm}{1.7em}
   & \lambda\,\int_{\partial \Omega}
   \rho_{\partial \Omega}(\vec{y})\,
   \Phi_n(\vec{x},\,\vec{y})\,d\/S
   = \dfrac{b(\vec{x})}{\langle \beta \rangle_{\Gamma}} -
   \lambda\,v_n(\vec{x}) \rule[-1.4em]{0mm}{1.2em}
  \end{split}
  &\mathrm{for}\;\; \vec{x} &\in \Gamma\/,
  \label{eq:wbie:in:dwng}\\
  \begin{split}
   -\rho_{\partial \Omega}(\vec{x}) &+
   2\,\int_{\Gamma}
   \rho_{\Gamma}(\vec{y})\,
   \Phi_m(\vec{x},\,\vec{y})\,d\/S\,+\\\rule{0mm}{1.7em}
   & 2\,\int_{\partial\Omega}
   \rho_{\partial \Omega}(\vec{y})\,
   \Phi_m(\vec{x},\,\vec{y})\,d\/S
   = 2\,(g_N(\vec{x}) - v_m(\vec{x}))
  \end{split}
  & \mathrm{for}\;\; \vec{x} &\in \partial\Omega\/.
  \label{eq:wbie:in:bc}
 \end{align}
\end{subequations}

\bigskip\noindent
Each of \eqref{eq:wbie:id} and \eqref{eq:wbie:in} is a
Fredholm integral equation of the second kind. This class of
integral equations has been studied extensively and there is
a number of well established numerical methods --- Boundary
Integral Methods (BIM) --- that can be used to solve for the
single and double layer densities ($\rho\/$ and
$\mu$)~\cite{mikhlin:57, atkinson:97}.
Nevertheless, although the densities can be computed
accurately and efficiently, the solution in $\Omega\/$
requires the computation of the single and double layer
potential solutions in~\eqref{eq:wbif}.
This last step can be
expensive when the solution is needed at a large number of
points.
In addition, the singularities in~\eqref{eq:wbif} make
evaluating the solution even more expensive in the vicinity
of $\Gamma\/$ and $\partial \Omega\/$.
These difficulties can be circumvented by computing the
solution inside $\Omega\/$ using finite differences, as
proposed by Mayo~\cite{mayo:84} and described
in~\S\ref{sub:solution:w2}.


%% file: 4solution43w2.tex
\subsection{Second solution component: finite differences
            computation}
\label{sub:solution:w2}
Once the single and double layer potential densities are
computed, \eqref{eq:wbif} can be used to extend $w\/$ to the
box $\mathcal{B}$ and obtain an equivalent definition of
$w\/$ as the solution to the following Laplace problem:
\begin{subequations}\label{eq:w2}
 \begin{align}
  \Delta\/w(\vec{x}) &= 0
  &\mathrm{for}\;\; \vec{x} &\in \mathcal{B}\/,
  \label{eq:w2:eq}\\     \rule{0mm}{1.1em}
  [w]_{\Gamma} &= 0
  &\mathrm{for}\;\; \vec{x} &\in \Gamma\/,
  \label{eq:w2:dwg}\\     \rule{0mm}{1.1em}
  [w_n]_{\Gamma} &= \phantom{-}\rho_{\Gamma}(\vec{x})
  &\mathrm{for}\;\; \vec{x} &\in \Gamma\/,
  \label{eq:w2:dwng}\\     \rule{0mm}{1.1em}
  [w]_{\partial \Omega} &= -\mu_{\partial \Omega}(\vec{x})
  &\mathrm{for}\;\; \vec{x} &\in \partial \Omega\/,
  \label{eq:w2:dwpo}\\     \rule{0mm}{1.1em}
  [w_n]_{\partial \Omega} &=
  \phantom{-}\rho_{\partial \Omega}(\vec{x})
  &\mathrm{for}\;\; \vec{x} &\in \partial \Omega\/,
  \label{eq:w2:dwnpo}\\
  \begin{split}
   w(\vec{x}) &=      \rule{0mm}{1.6em}
   \int_{\Gamma}
   \rho_{\Gamma}(\vec{y})\,\Phi(\vec{x},\,\vec{y})\,d\/S \\
   &+ \rule{0mm}{1.6em} \int_{\partial \Omega}
   \bigl(\rho_{\partial \Omega}(\vec{y})\,
   \Phi(\vec{x},\,\vec{y}) +
   \mu_{\partial \Omega}(\vec{y})\,
   \Phi_m(\vec{x},\,\vec{y})\bigr)
  \end{split}
  &\mathrm{for}\;\; \vec{x} &\in \partial\mathcal{B}\/.
  \label{eq:w2:bc}
 \end{align}
\end{subequations}
This problem is, again, of the type that can be solved using
the CFM. Furthermore, it is also a problem in the box
$\mathcal{B}\/$, so that the discretized system can be
inverted very efficiently using the FFT.

Once $v\/$ and $w\/$ have been computed as the solutions to
\eqref{eq:v} and \eqref{eq:w2}, respectively, the solution
to (\ref{eq:poisson}--\ref{eq:bc}) follows from $u = v +
w\/$.
%

%% file: 4solution44open.tex
\subsection{Open space problems}
\label{sub:solution:open}
The algorithm described in
\S\ref{sub:solution:v}--\S\ref{sub:solution:w2} can be
extended to open space problems easily. The adaptations
required are described below.

In the computation of the first solution component, equation
\eqref{eq:v}, $\Omega\/$ is replaced by the support of
$f\/$, denoted by $\Omega_f\/$. In this case, the box
$\mathcal{B}\/$ must enclose $\Omega_f\/$. Then, the first
solution component is set to be
\begin{align}\label{eq:vo}
 v(\vec{x}) &= 
 \begin{cases}
  0\\ \rule{0mm}{1.1em}
  \text{solution to \eqref{eq:v}}
 \end{cases}&
 \begin{matrix}
  \mathrm{for}\;\;\vec{x} \\ \rule{0mm}{1.1em}
  \mathrm{for}\;\;\vec{x}
 \end{matrix}&
 \begin{array}{l}
  \in \mathbb{R}^{\nu}/\Omega_f\/,\\ \rule{0mm}{1.1em}
  \in \Omega_f\/.
 \end{array}
\end{align}
Equation~\eqref{eq:vo} results in jumps on $v\/$ and $v_n\/$
across $\partial \Omega_f\/$.
These jumps are accounted for by the second solution
component, which is required to satisfy:
\begin{subequations}\label{eq:w1:o}
 \begin{align}
  [w]_{\partial \Omega_f} &= -[v]_{\partial \Omega_f}\/,
  \label{eq:w1:o:dwpo}\\ \rule{0mm}{1.1em}
  [w_m]_{\partial \Omega_f} &= -[v_m]_{\partial \Omega_f}\/.
  \label{eq:w1:o:dwnpo}
 \end{align}
\end{subequations}
\revc{Furthermore, since $w$ and $u$ are equal outside
$\Omega_f\/$, the asymptotic behavior of $w\/$ is given by
\begin{subequations}\label{eq:w1:o:infty}
 \begin{align}
  w(\vec{x})
  &\; \sim \; \dfrac{F}{2\pi}\ln\lvert\vec{x}\rvert,
  &\mbox{for $\lvert\vec{x}\rvert \gg 1$ and $\nu=2$,}\\
  w(\vec{x})
  &\; \sim \;
   \dfrac{F}{(\nu-2)A_{\nu}\lvert\vec{x}\rvert^{\nu-2}},
  &\mbox{for $\lvert\vec{x}\rvert \gg 1$ and $\nu=3$,}
 \end{align}
\end{subequations}
where $A_{\nu}\/$ is  area of the unit sphere in
$\mathbb{R}^\nu\/$ and $F\/$ is defined in \eqref{eq:F}.}
As a consequence, the second solution component can be
represented by the integral formulation
\begin{align}\label{eq:wbif:o}
 \begin{split}
  w(\vec{x}) &=
  \int_{\Gamma}
  \rho_{\Gamma}(\vec{y})\,\Phi(\vec{x},\,\vec{y})\,d\/S \\
  &-\int_{\partial \Omega_f}
  \bigl([v_m]_{\partial \Omega_f}(\vec{y})\,
  \Phi(\vec{x},\,\vec{y}) +
  [v]_{\partial \Omega_f}(\vec{y})\,
  \Phi_m(\vec{x},\,\vec{y})\bigr)\,d\/S
  \rule{0mm}{1.8em}
 \end{split}
 &\mathrm{for}\;\; \vec{x} &\in \mathbb{R}^{\nu}\/,
\end{align}
where $\rho_{\Gamma}$ is the solution to the following
integral equation:
\begin{align}\label{eq:wbie:o}
 \rho_{\Gamma}(\vec{x}) &+
 \lambda\,\int_{\Gamma}
 \rho_{\Gamma}(\vec{y})\,\Phi_n(\vec{x},\,\vec{y})\,d\/S
 = \dfrac{b(\vec{x})}{\langle \beta \rangle_{\Gamma}} -
 \lambda\,v_n(\vec{x})
 &\mathrm{for}\;\; \vec{x} &\in \Gamma\/.
\end{align}

The remainder of the algorithm is identical to the case of
an interior Poisson problem, described
in~\S\ref{sub:solution:v}--\S\ref{sub:solution:w2}.


%% file: 4solution45implementation.tex
\subsection{Implementation details}
\label{sub:solution:implementation}
We complete the description of the algorithm with some
implementation details. First, the BIM requires a
discretization of $\Gamma\/$ and $\partial \Omega\/$.
However, these discretizations are completely independent of
the computational grid used by the finite differences in
\S\ref{sub:solution:v} and \S\ref{sub:solution:w2}. In fact,
it is possible to have a BIM which is more accurate than the
finite differences scheme. In this case the discretization
used for $\Gamma\/$ and $\partial \Omega\/$ can be coarser
than the finite differences grid.

\revc{Second, in general the boundary integral formulation
results in non-symmetric equations ---
see~\eqref{eq:wbie:id}, \eqref{eq:wbie:in},
and~\eqref{eq:wbie:o}.
This lack of symmetry stems from $\Phi_n$ and $\Phi_m$,
which are not symmetric functions\footnote{\revc{We
call a function $f$ of two variables symmetric when
$f(x_1,x_2) = f(x_2,x_1)\/$.}}.
As a consequence, the linear systems that result from the
discretization of these equations are also not symmetric,
and this fact must be taken into consideration when solving
these linear systems.
Here we adopt the Generalized Minimal Residual (GMRES)
method to solve these linear systems, since it has been
identified as an efficient method to solve integral
equations~\cite{mayo:jssc:1992}.
The only exception is the case of poorly conditioned
problems described in~\S\ref{sub:solution:poorcond}.}

Third, the rectangular domain $\mathcal{B}\/$ is arbitrary.
However, to reduce the number of grid points outside the
region of interest, $\mathcal{B}\/$ should enclose
$\Omega\/$ as tightly as possible.
On the other hand, evaluating \eqref{eq:w2:bc} too close to
$\partial \Omega\/$ is difficult.
Hence, the distance from $\partial \Omega\/$ to
$\partial \mathcal{B}\/$ should not be too small.
In our calculations we used the distance of three times the
discretization spacing used for $\partial \Omega\/$, as
suggested by Mayo~\cite{mayo:84}.

Fourth, the CFM demands the computation of the jump
conditions at a number of points along the interfaces.
In the present algorithm, the jump conditions for $w\/$ are
given in terms of the potential densities computed using a
BIM.
Hence, interpolation is needed to compute the values at the
points required by the CFM.

Fifth, the algorithm requires the evaluation of the normal
derivative $v_n\/$ along $\Gamma\/$.
There is not a unique way to compute $v_n\/$.
The solution adopted here is to use Hermite polynomials to
represent $v\/$ in grid cells that are crossed by
$\Gamma\/$.
This approach requires the computation of derivatives of
$v\/$ at grid nodes adjacent to $\Gamma\/$.
We perform these computations with finite differences, using
the correction function to extend the solution across
$\Gamma\/$.
Furthermore, although $v\/$ is discontinuous, $v_n\/$ is
continuous across $\Gamma\/$ --- see \eqref{eq:v:dung}.
Thus $v_n\/$ can be computed using the Hermite polynomial
representation of either $v^+\/$ or $v^-\/$.
The same argument is valid for the computation of $v_m\/$
along $\partial \Omega\/$.

\reva{Finally, in principle, the algorithm can solve problems
that involve interfaces with multiple corners.
In this case, the user must choose a BIM that produces
accurate solutions in the presence of corners.
Some options are described in~\cite{atkinson:97}.
Also, the implementation of the CFM described
in~\cite{marques:11} needs to be adaptated to take into
account the presence of corners.
In this paper we restrict our attention to smooth interfaces
to simplify the presentation of the algorithm.}
%

%% file: 4solution46singular.tex
\subsection{Singular problems}
\label{sub:solution:singular}
As mentioned in remark~\ref{rmk:problem:neumann}, the
solution to the interior Neumann problem is only defined up
to a constant. In this section we discuss the use of the
Generalized Minimal Residual (GMRES)
method~\cite{trefthen:97} to circumvent this issue.

Consider a general linear system of the form
\begin{equation}\label{eq:ls}
 Ax = y\/.
\end{equation}
\rrrc{The following result follows from theorem 2.6 of
ref.~\cite{brown:97}}
%
\begin{cor}\label{cor:gmres}
 GMRES is guaranteed to produce a solution to \eqref{eq:ls}
 without breakdown if
 \begin{enumerate}[(i)]
  \item $\mathrm{range}(A)\cap\mathrm{nullspace}(A) =
   \emptyset\/$, and
  \item $y \in \mathrm{range}(A)\/$.
 \end{enumerate}
 In addition, GMRES produces the solution that minimizes the
 least-squares problem $\lVert Ax-b \rVert_2$.\myremarkend
\end{cor}

For the sake of clarity, in this section we limit our
discussion to the interior Neumann problem without jumps ---
the same arguments apply to the full problem. In this
situation, \eqref{eq:wbie:in} reduces to 
\begin{equation}\label{eq:K}
 \begin{split}
  \mathcal{K}\rho_{\partial \Omega} &=
  -\rho_{\partial \Omega}(\vec{x}) +
  2\,\int_{\partial \Omega}
  \rho_{\partial \Omega}(\vec{y})\,
  \Phi_m(\vec{x},\,\vec{y})\,d\/S\\ \rule{0mm}{1.2em}
  &= 2\,(g_N(\vec{x}) - v_m(\vec{x}))
 \end{split}
 \qquad\mathrm{for}\;\; \vec{x} \in \partial \Omega\/,
\end{equation}
where $\mathcal{K}\/$ denotes the integral operator
associated with the interior Neumann problem. Below we use
known properties of $\mathcal{K}\/$ to show that
\eqref{eq:K} satisfies conditions (i) and (ii) of
corollary~\ref{cor:gmres} at the continuous level. While we
cannot prove it, our numerical experiments indicate that
these properties carry over to the discretizations that we
used, and that GMRES is very robust in solving the linear
systems that arise.

\medskip\noindent
\emph{Proof that \eqref{eq:K} satisfies (i) and
                 (ii) of corollary~\ref{cor:gmres}.}

\smallskip
First: it is known (see ref.~\cite{atkinson:97}) that
$\mathrm{nullspace}(\mathcal{K})\/$ is the one-dimensional
space spanned by a $\psi\/ \in C^1(\partial \Omega)\/$ with
the property
\begin{equation}\label{eq:psi}
 \int_{\partial \Omega} \psi\Phi\,d\/S = 1\/.
\end{equation}
Furthermore, as shown in ref.~\cite{pogorzelski:66}, it
follows from the maximum/minimum principle of the Laplace
equation that either $\psi \le 0\/$ or $\psi \ge 0\/$. As a
consequence, 
\begin{equation}\label{eq:null}
 \int_{\partial \Omega} \psi\,d\/S \ne 0\/.
\end{equation}

Second: it also known (see ref.~\cite{atkinson:97}) that
\begin{equation}\label{eq:range}
 \int_{\partial \Omega} \varphi\,d\/S = 0
 \Longleftrightarrow \varphi \in
 \mathrm{range}(\mathcal{K})
 \quad \forall \varphi \in C^1(\partial\Omega)\/.
\end{equation}
Thus, it follows from \eqref{eq:null} and \eqref{eq:range}
that
\begin{equation}\label{eq:rnK}
 \mathrm{range}(\mathcal{K}) \cap
 \mathrm{nullspace}(\mathcal{K}) = \emptyset.
\end{equation}

Finally: it follows from the divergence theorem that
\begin{equation}\label{eq:div}
 \int_{\partial \Omega} g_N\,d\/S = 
 \int_{\partial \Omega} v_m\,d\/S = 
 \dfrac{1}{\beta}\int_{\Omega} f\,d\/V\/,
\end{equation}
where it is assumed that $\beta = \beta^+ = \beta^-\/$.
As a consequence,
\begin{equation}\label{eq:b}
 (g_N - v_m) \in \textrm{range}(\mathcal{K})\/.
\end{equation}
Therefore, \eqref{eq:K} satisfies conditions (i) and (ii) of
corollary~\ref{cor:gmres} at the continuous level.\myremarkend

Let $Ax = b\/$ be the linear system resulting from one of
our discretizations of \eqref{eq:K}. Our numerical
experiments show that GMRES is very robust in solving
$PAx = Pb\/$, where
\begin{equation}\label{eq:P}
  P = I - eq^T,
\end{equation}
$e = \{1 \ldots 1\}^T$, and $q\/$ is the vector of
quadrature weights used to discretize the integral operator
in \eqref{eq:K}. The projection operator $P\/$ guarantees
that, despite discretization errors, (i) $Pb \in
\mathrm{range}(PA)\/$, and (ii) that $\mathrm{range}(PA)\/$
lies in the space of zero-mean vectors. These conditions are
valid for the continuous operator $\mathcal{K}\/$, and are
essential to show that the conditions of
corollary~\ref{cor:gmres} are satisfied at the continuous
level. Furthermore, by using the same quadrature rule to
construct the projection $P\/$, this step is guaranteed not
affect the accuracy order of the numerical approximation.
%

%% file: 4solution47poorcond.tex
\subsection{Poorly conditioned problems}
\label{sub:solution:poorcond}
As explained in \S\ref{sub:problem:poorcond}, Poisson
problems that involve jumps in $[\beta u_n]\/$ become poorly
conditioned when $\beta^- \gg \beta^+\/$. In this section we
discuss how this issue affects the algorithm presented here,
and propose a solution. For the sake of clarity, we limit
this discussion to open space problems. However, the same
arguments apply to interior problems.

When $\beta^- \gg \beta^+\/$, \eqref{eq:wbie:o} is poorly
conditioned because it approaches the singular case of a
Neumann problem.
In this case, although \eqref{eq:wbie:o} is poorly
conditioned, the problem is not singular.
Hence, corollary~\ref{cor:gmres} guarantees that GMRES will
produce a solution without breakdown.
However, the poor conditioning of \eqref{eq:wbie:o} means
that the solution is very sensitive to truncation errors
that occur in finite accuracy arithmetic.
Our examples show that, depending on the ratio
$\beta^-/\beta^+\/$, this effect may be large enough to
render the solution unreliable, even when very accurate
numerical methods are used.

\revc{We address this issue by add adding to the equations a
constraint that removes the poor conditioning.
This constraint must be based on an identity that is
satisfied by the solution to the Poisson equation.
For instance, the following identity follows from
\eqref{eq:poisson}, \eqref{eq:jumpcond:dun}, and the
divergence theorem,
\begin{equation}\label{eq:intdun}
 \int_{\Gamma} [u_n]_{\Gamma}\,d\/S = 
 \dfrac{2}{\lambda + 2}
 \Bigl(\dfrac{1}{\langle\beta\rangle_{\Gamma}}
 \int_{\Gamma} b\,d\/S 
 -\dfrac{\lambda}{\beta^-}\int_{\Omega^-} f^-\,d\/V\Bigr)\/.
\end{equation}
Then, from \eqref{eq:sl:jumps:un} and \eqref{eq:intdun}, it
can be shown that the solution must satisfy the following
constraint:
\begin{equation}\label{eq:intrho}
 \begin{split}
  \int_{\Gamma} \rho_{\Gamma}\,d\/S
  &= \int_{\Gamma} [u_n]\,d\/S\\ \rule{0mm}{1.8em}
  &= \dfrac{2}{\lambda + 2}
     \Bigl(\dfrac{1}{\langle\beta\rangle_{\Gamma}}
     \int_{\Gamma} b\,d\/S 
   - \dfrac{\lambda}{\beta^-}
     \int_{\Omega^-} f^-\,d\/V\Bigr)\/.
  \end{split}
\end{equation}
We also notice that, in the limit when $\beta^-/\beta^+ \to
\infty\/$ ($\lambda \to -2\/$) this constraint becomes an
equation that removes the singular nature of the Neumann
limit, by selecting a specific solution out of the
infinitely many possible. Thus we add this extra constraint
to the system to be solved, to remove the poor conditioning
of the $\beta^- \gg \beta^+$ problem. Our numerical
calculations indicate that this works.}

Furthermore, the right hand side of \eqref{eq:intrho}
depends on known problem data only, and should be
precomputed, so that \emph{the actual condition imposed
is: $ \int_{\Gamma} \rho_{\Gamma}\,d\/S =\/$ known
constant.} The poor problem conditioning is then shifted to
the accurate computation of the right hand side in
\eqref{eq:intrho}, which is done separately. When the
denominator $(\lambda + 2)\/$ is small, the integrals in the
numerator must be computed very carefully, so that their
difference is known accurately.

\revc{The accuracy with which the right hand side in
\eqref{eq:intrho} needs to be computed depends on the
accuracy desired for the underlying Poisson problem.
An error of size $\varepsilon\/$ in the computation of
\eqref{eq:intrho} results in a shift of $\rho_{\Gamma}\/$ by
$\varepsilon/S_{\Gamma}$, where $S_{\Gamma}\/$ denotes the
area of the surface $\Gamma\/$.
In turn, this error propagates to $u$ via the jump
condition~\eqref{eq:w2:dwng} and the boundary
condition~\eqref{eq:w2:bc}. The error in $u$ depends on the
geometry of $\Gamma\/$ and $\partial \Omega\/$, but it
scales as $\ord{\varepsilon}$.}

In the examples shown
in~\S\ref{sec:results}, the integrals are computed with
spectral accuracy using the trapezoidal
rule
\footnote{\reva{Consider
a $2\pi$-periodic function $\phi \in H^q(2\pi)$,
$q > 1/2$. Then, lemma 7.3.3 of ref.~\cite{atkinson:97}
shows that $|\int_0^{2\pi}\phi(s) ds - T_h(\phi)| \le \,
C||\phi||_q \, h^q$, where $T_h$ denotes the trapezoidal
rule with discretization size $h$, and
$C$ is a constant (independent of $\phi$, $q$, and $h$).}}.
In real applications, physical reasoning may help to deduce
the correct value for the integral of $\rho_{\Gamma}\/$. For
instance, in many fluid flow problems the integral is
identically zero.

Finally, when \eqref{eq:intrho} is enforced as an additional
condition to \eqref{eq:wbie:o}, the result is an
overdetermined linear system.
\revc{We solve the normal equation associated with this
overdetermined system using the Conjugate Gradient (CG)
method~\cite{trefthen:97}.}


%% file: 4solution48accuracy.tex
\subsection{Accuracy and efficiency}
\label{sub:solution:accuracy}
In this section we discuss the accuracy and efficiency of
the algorithm. The accuracy is determined by five factors:
\begin{enumerate}
 \item \textbf{Representation of interfaces and boundaries.}
  The interface and boundary conditions must be enforced
  with accuracy comparable with the rest of the algorithm.
  Thus, the position of the interfaces and boundaries must
  be known in an equally accurate fashion. For the examples
  in~\S\ref{sec:results} we represent surfaces with equally
  spaced markers. Geometrical properties are
  computed with trigonometric
  interpolation~\cite{bulirsch:02}.
 \item \textbf{Accuracy of the BIM.}
  The accuracy of these methods depends on the smoothness
  of: the interfaces, the boundaries, and the data provided
  on these surfaces. For smooth and well resolved
  surfaces, Nystrom's method~\cite{atkinson:97} is
  guaranteed to converge as fast as the quadrature rule
  used to approximate the integrals.
 \item \textbf{Interpolation of the single and double layer
  potential densities.} As pointed out
  in~\S\ref{sub:solution:implementation}, interpolation of
  the solution obtained with the BIM, to the points required
  by the CFM, may be needed. In this paper the examples are
  in 2-D, with smooth geometries.  Thus we use trigonometric
  interpolation~\cite{bulirsch:02}, which can be efficiently
  computed using the FFT, and has optimal accuracy.
 \item \textbf{Accuracy of the CFM.}
  There is no ``in principle'' limit to the CFM
  accuracy~\cite{marques:11}, provided that the data
  (source terms and jump functions) are smooth enough. In
  our examples we use a \fth\ order implementation.
 \item \textbf{Computation of normal derivatives.}
  When the problem involves discontinuity interfaces, or
  Neumann boundary condition, the algorithm requires the
  computation of the normal derivative of $v\/$ along
  $\Gamma\/$ or $\partial \Omega\/$. Within the context of
  our implementation, this computation causes an order loss
  in the CFM, relative to its nominal accuracy. The reason
  is that the Hermite interpolants used to calculate the
  correction function produce derivatives with one order
  less accuracy than the function values.
\end{enumerate}
The overall accuracy of the algorithm is determined by the
least accurate of the factors listed above. Since, in
principle, each of these factors can be made as accurate as
needed; there is no inherent limit to the algorithm order.

In the examples in~\S\ref{sec:results}, we use a \fth\ order
implementation of the CFM. Hence, the overall accuracy of
the algorithm is (i) \fth\ order in the case of an interior
Dirichlet problem without a discontinuity interface, and
(ii) \trd\ order in cases involving either Neumann boundary
conditions or discontinuity interfaces, since these problems
require the computation of derivatives of $v\/$.

For the purpose of evaluating the operation count of the
algorithm, let us assume that (i) the BIM interfaces and
boundaries are discretized using a total of $k\/$ nodes, and
(ii) the finite differences discretization of the
computational domain involves $M\/$ nodes.\footnote{These
   two discretizations are independent of each other.}
Then the break down of the operation count is as follows
(note that the space dimension is $\nu\/$).
\begin{enumerate}
 \item \textbf{BIM operation count.}
  This depends on the specific choice of
  method. In principle, a general BIM requires $\ord{k^3}\/$
  operations. However, there is a number of techniques that
  can be used to reduce the operation count to $\ord{k^2}\/$
  or even to $\ord{k}\/$, see~\cite{rokhlin:85, canning:92,
  nabors:94}.
 \item \textbf{Operation count of computing boundary
               conditions.}
  Equation \eqref{eq:w2:bc} requires integrations to
  evaluate the boundary conditions. This yields $\ord{k}\/$
  operations per node on the boundary. Thus the total
  operation count from the boundary conditions is
  $\ord{kM^{(\nu\/-1)/\nu\/}}\/$.
 \item \textbf{Operation count of interpolation.}
  Depending on the technique used, the operation count of
  computing interpolants for the potential densities varies
  between $\ord{k}\/$ and $\ord{k^2}\/$. After the
  interpolants are known, the operation count of evaluating
  the densities at the locations needed by the CFM is
  $\ord{M^{(\nu\/-1)/\nu\/}}\/$.
 \item \textbf{CFM Operation count.}
  Computing the correction function requires the solution of
  a small linear system ($12\times\/12\/$ for \fth\ order
  accuracy in 2-D) at each grid node close to the interface
  or the boundary. The operation count of this step is
  $\ord{M^{(\nu\/-1)/\nu\/}}\/$. However, note that these
  linear systems depend on the geometry of the problem only
  --- thus their coefficient matrices need to be computed
  only once. As a consequence, even though the CFM is used
  more than once to obtain the full solution, the additional
  operation count incurred over a single use is rather
  minimal.
 \item \textbf{Operation count of finite differences
               discretization.}
  Since all the problems solved with finite differences are
  defined in a rectangular domain, the resulting linear
  systems can be inverted using the FFT. This is one of the
  fastest methods available to solve the Poisson equation,
  with operation count $\ord{M\log\/M}\/$.
\end{enumerate}


%% file: 5results.tex
\section{Results}\label{sec:results}
Here we present three examples of computations in 2-D using
the algorithm proposed in~\S\ref{sec:solution}. In the first
example we consider the problem of imposing Dirichlet or
Neumann conditions on an immersed boundary --- \ie\ interior
Dirichlet or Neumann problems with no interfaces. In the
second example we solve an open space Poisson problem, 
including a discontinuity interface. Finally, in the third
example we consider a combination of the previous two:
interior Poisson problems with Dirichlet or Neumann 
conditions on an immersed boundary, and a discontinuity
interface. Furthermore, in the second and third examples we
consider very large $\beta\/$ ratios of $10^6\/$.

The algorithm presented in~\S\ref{sec:solution} combines
different numerical methods, each of which has several
possible variants. The specific variants used for the
calculations presented in this section are as follows.
%
\begin{itemize}
 \item For the boundary integral equation we use Nystrom's
  method with trapezoidal quadrature rule~\cite{mayo:84,
  atkinson:97, rokhlin:85}.
  For smooth interfaces and data, this method has optimal
  convergence and accuracy.
  \revc{Furthermore, in most cases we solve the linear system
  that results from this discretization using the General
  Minimal Residual (GMRES) method~\cite{trefthen:97}.
  The exceptions are cases in which $\beta^- \gg \beta^+\/$.
  In these cases we add the constraint~\eqref{eq:intrho},
  resulting in overdetermined linear systems.
  We solve the normal equations related to these
  overdetermined linear systems using the Conjugate Gradient
  (CG) method~\cite{trefthen:97}.}
 \item For the interpolation of single and double layer
  potential densities we use
  FFT-based trigonometric interpolants~\cite{bulirsch:02}.
 \item Interfaces and boundaries immersed in regular
  Cartesian grids are represented by
  \rrrc{markers equispaced along the
  curves.}\footnote{\rrrc{For 3-D versions of the algorithm
  see the last paragraph in section~\ref{sec:conclusion}.}}
  Geometrical information is computed based on trigonometric
  interpolation.
 \item The Laplace and Poisson equations are discretized
  using the standard 9-point stencil~\cite{marques:11}. This
  discretization results in \fth\ order accuracy.
 \item We use the \fth\ order implementation of the CFM
  described in ref.~\cite{marques:11}.
\end{itemize}
%
Nystrom's method converges rapidly for problems involving
smooth data, such as the examples below. For this reason, we
were able to obtain satisfactory results by setting
\begin{equation*}
 h_{\Gamma} \approx h_{\partial \Omega} \approx 2h\/,
\end{equation*}
where $h_{\Gamma}\/$ and $h_{\partial \Omega}$ denote
the discretization spacing along $\Gamma\/$ and $\partial
\Omega\/$ used in Nystrom's method, and $h\/$ denotes the
spacing of the grid used in the finite differences steps.
The relationships between $h_{\Gamma}\/$, $h_{\partial \Omega}$,
and $h\/$ are approximate because Nystrom's method requires
a uniform discretization.

In the three examples considered here, the boundaries and
interfaces are either circles, or ``smooth five-pointed
stars''. These geometries can be parametrized as follows:
\begin{align}\label{eq:star}
 x = R\cos(\theta), && y = R\sin(\theta),
\end{align}
where $R = r_0 + \delta\,\sin\/(5\,\theta)\/$.
Here $\delta\/$ and $r_0\/$ are parameters defined in each
example.
\reva{Note that with the parametrization~\eqref{eq:star}, an
equi-spaced set of markers along the boundaries/interfaces
does not correspond to uniformly spaced values of $\theta$.
Thus, in order to determine the $\theta$-coordinates of the
markers, a Newton iteration is used in a pre-processing
step. This is explained below.

First, Gaussian quadrature is applied to a very fine
discretization of the boundary/interface to compute its
length.
Once the length of the boundary/interface is known, the
spacing that corresponds to a distribution of uniformly
distanced markers ($h_{\partial \Gamma}\/$ or
$h_{\Gamma}\/$) can be determined.
Next, the iterative process begins.
A distribution of markers that correspond to uniformly
spaced values of $\theta\/$ is used as an initial guess.
Then, for each new marker, a Newton iteration is used to
solve the problem
\begin{equation}
 F(\vec{x}_i) =
 \text{dist}(\vec{x}_i,\vec{x}_{i-1}) 
 - h_{\partial \Omega} = 0
 \qquad \mbox{(or $h_{\Gamma}\/$)},
\end{equation}
where the distance between the current guess and the
previous marker is computed using Gaussian quadrature.
Finally, the process is repeated until the position of all
markers has been determined.}

\subsection{Example 1. Interior Dirichlet and Neumann
            problems}
\label{sub:results:interior}
\begin{figure}[htb!]
 \begin{center}
  \includegraphics[width=2.6in]
  {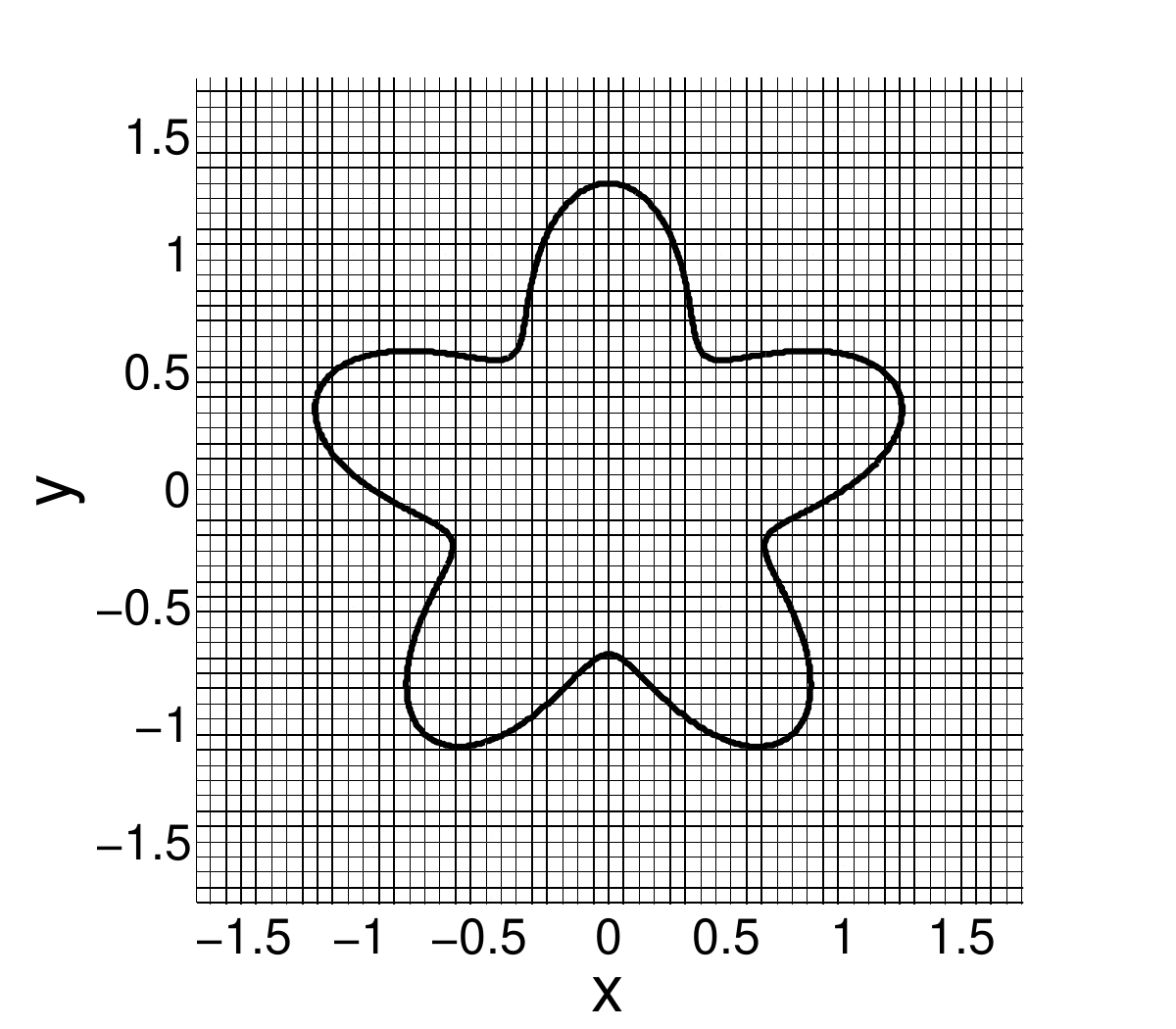}
  \includegraphics[width=2.6in]
  {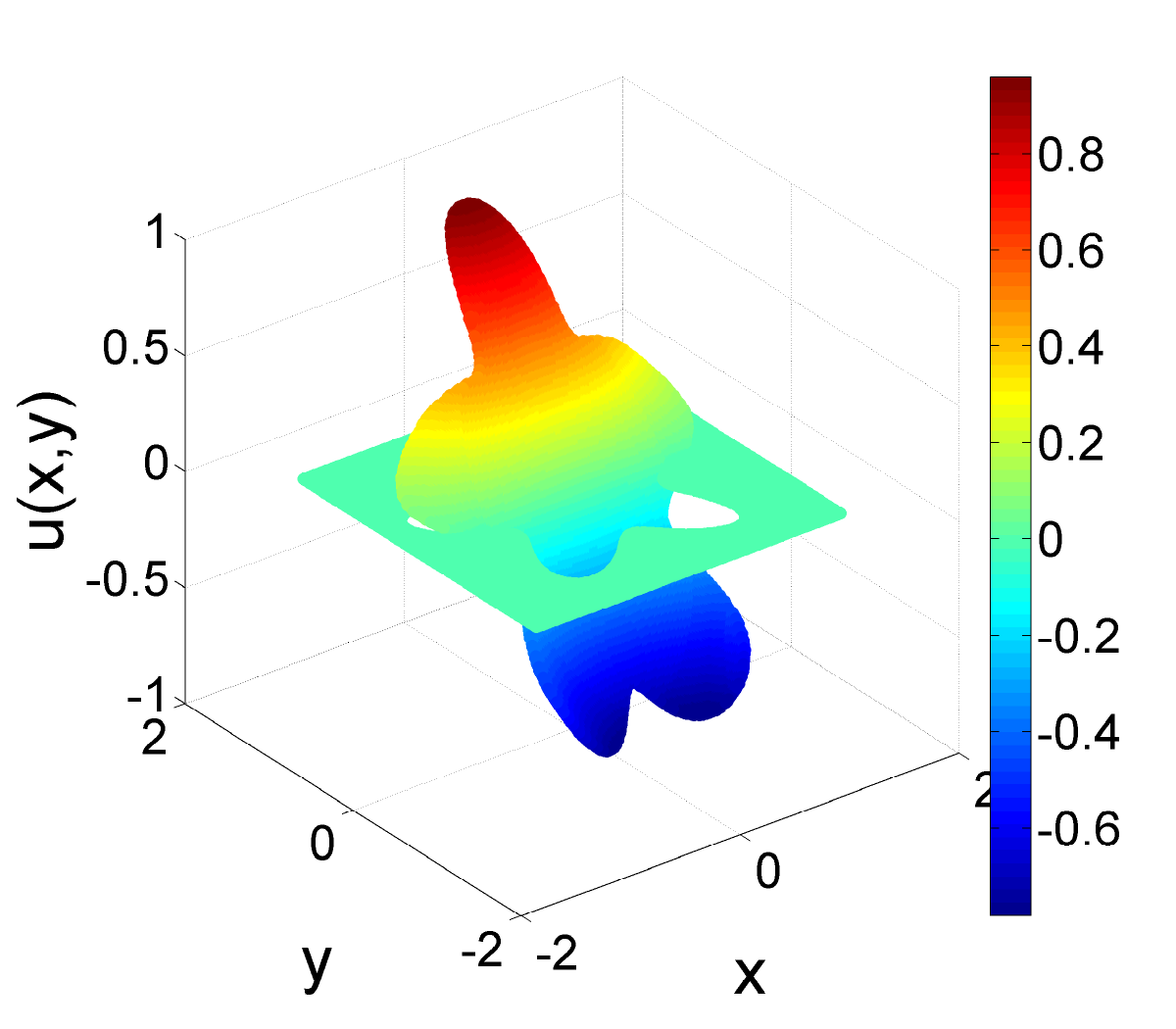}\\
 \end{center}
 \caption{Left: solution domain $\Omega\/$ for example 1,
    embedded into a Cartesian grid. Right: solution
    obtained with the algorithm proposed in this paper.}
 \label{fig:ex1:grid}
\end{figure}

In this example, the solution domain is the interior region
bounded by the curve defined by \eqref{eq:star}, with
$\delta = 0.3\/$ and $r_0 = 1\/$. Figure~\ref{fig:ex1:grid}
shows $\Omega\/$ immersed in a regular Cartesian
discretization of a square box.

For this domain, consider the Poisson equation associated
with the exact solution
\begin{equation}\label{eq:ex1}
 u(x,y) = \cos\/(x)\sin\/(y)\/,
\end{equation}
with either Dirichlet or Neumann boundary condition.
Figure~\ref{fig:ex1:grid} shows a plot the solution
obtained with the present algorithm in the Dirichlet case
(the solution outside $\Omega\/$ is set to zero). In the
Neumann problem, the solution is only defined up to an
arbitrary constant, and the constant ``picked'' by GMRES may
not correspond to \eqref{eq:ex1}. Hence we shift the Neumann
solution such that $u(0,0) = 0\/.$ After this step, both
Dirichlet and Neumann solutions become visually
indistinguishable.

\begin{figure}[htb!]
 \begin{center}
  \includegraphics[width=2.6in]
  {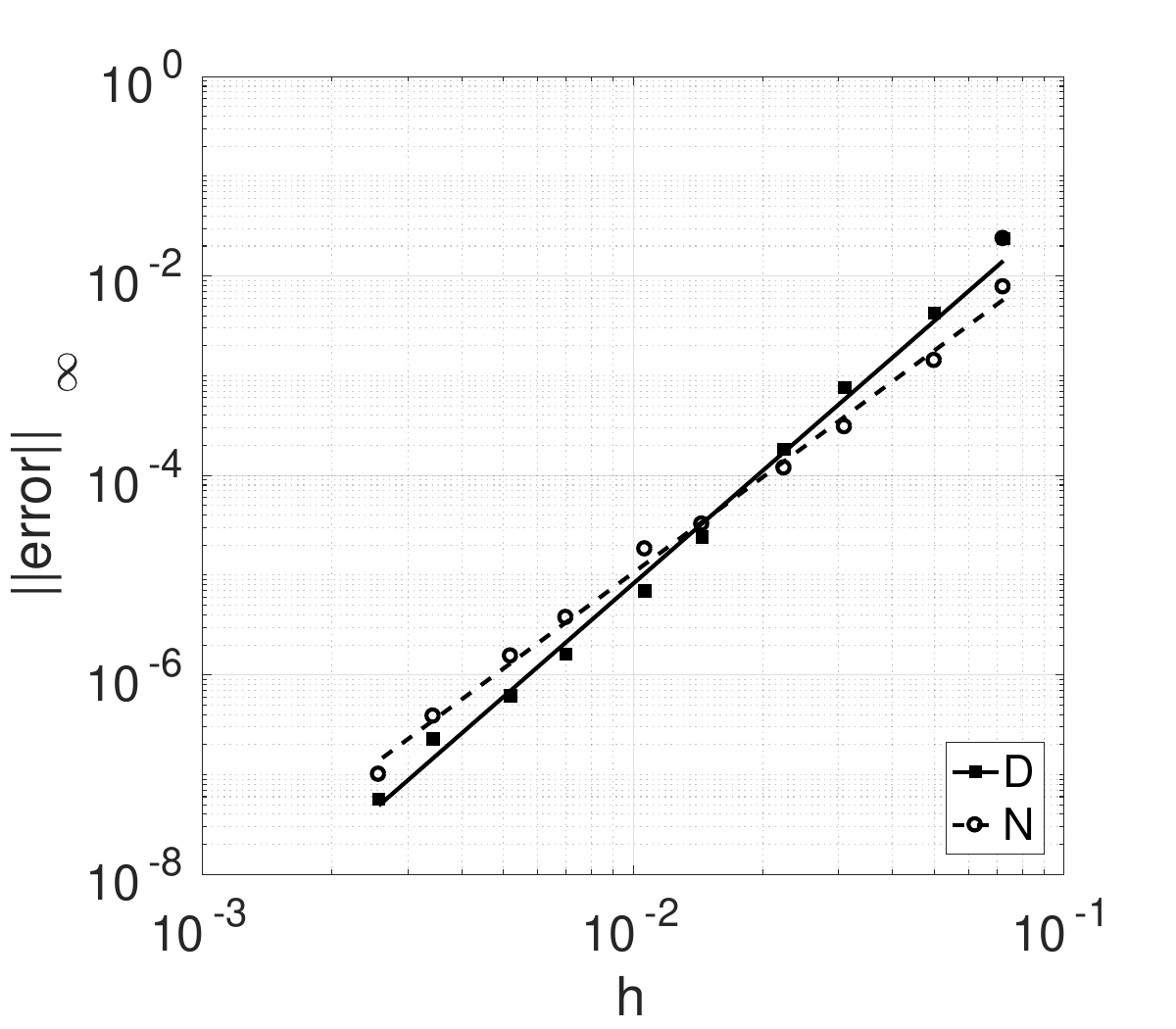}
  \includegraphics[width=2.6in]
  {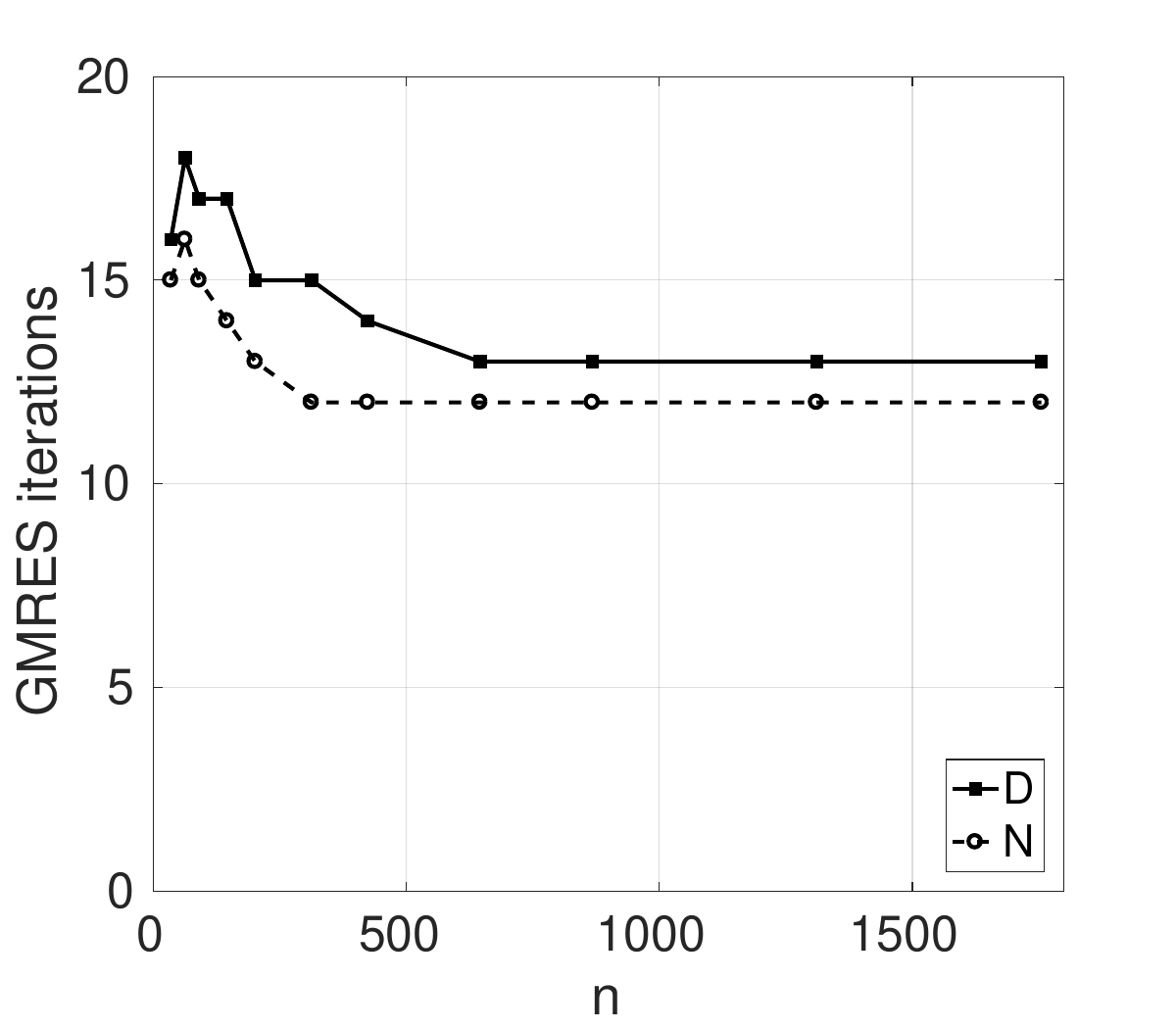}
 \end{center}
 \caption{\revc{Left: error convergence for the Dirichlet
    and Neumann Poisson problems in example 1.
    Error is measured in the $L_{\infty}\/$ norm.
    The straight lines were obtained by least squares fits
    and their slopes correspond to the measured convergence
    rates -- see table~\ref{tab:convergence}.
    Right: number of GMRES iterations needed to compute the
    potential distributions at the boundary
    $\partial \Omega\/$.
    Here $n$ indicates the number of markers used to discretize
    the boundary.}}
 \label{fig:ex1:convergence}
\end{figure}

Figure~\ref{fig:ex1:convergence} illustrates the convergence
of the method with a plot of the $L_{\infty}\/$ norm of the
error versus $h\/$. The measured convergence rates are
listed in table~\ref{tab:convergence} at the end of this
section.
The measured rates indicate that the observed convergence is
consistent with \fth\ order in the Dirichlet case, and \trd\
order in the Neumann case.
These results are as expected since we apply the algorithm
with a combination of \fth\ order numerical methods. The
order loss in the Neumann problem occurs because the
algorithm requires the computation of $v_m\/$ along
$\partial \Omega\/$.

\revc{Figure~\ref{fig:ex1:convergence} also shows the number
of GMRES iterations needed to compute the potential
distribution of the boundary integral formulation.
No preconditioner is used, and the residual tolerance is set
to $10^{-10}$.
This figure shows that GMRES converges with a relatively
small number of iterations, and that the number of
iterations that does not depend on the size of the problem.
These observations indicate that GMRES is an adequate solver
for this problem.}

\subsection{Example 2. Open space problem with a
            discontinuity interface} \label{sub:results:open}
In this example we solve the Poisson problem, defined in
$\mathbb{R}^2\/$, associated with the exact solution
\begin{subequations}\label{eq:ex2}
 \begin{align}
   u^+(x,\,y) &= \exp(-(x^2+y^2)/0.0008)\/,\\
                 \rule{0mm}{1.1em}
   u^-(x,\,y) &= \cos\/(x)\sin\/(y) + 1.5\/.
 \end{align}
\end{subequations}
The discontinuity interface, $\Gamma\/$, is defined by the
curve \eqref{eq:star}, with parameters $\delta = 0.03\/$ and
$r_0 = 0.1\/$. Furthermore, we consider two cases in which
the coefficients $\beta^{\pm}\/$ have large jumps across the
discontinuity interface:
%
\begin{enumerate}[(i)]
 \item $\beta^- = 1\/$, $\beta^+ = 10^6\/$,
 \item $\beta^- = 10^6\/$, $\beta^+ = 1$.
\end{enumerate}

\begin{figure}[htb!]
 \begin{center}
  \includegraphics[width=2.6in]
  {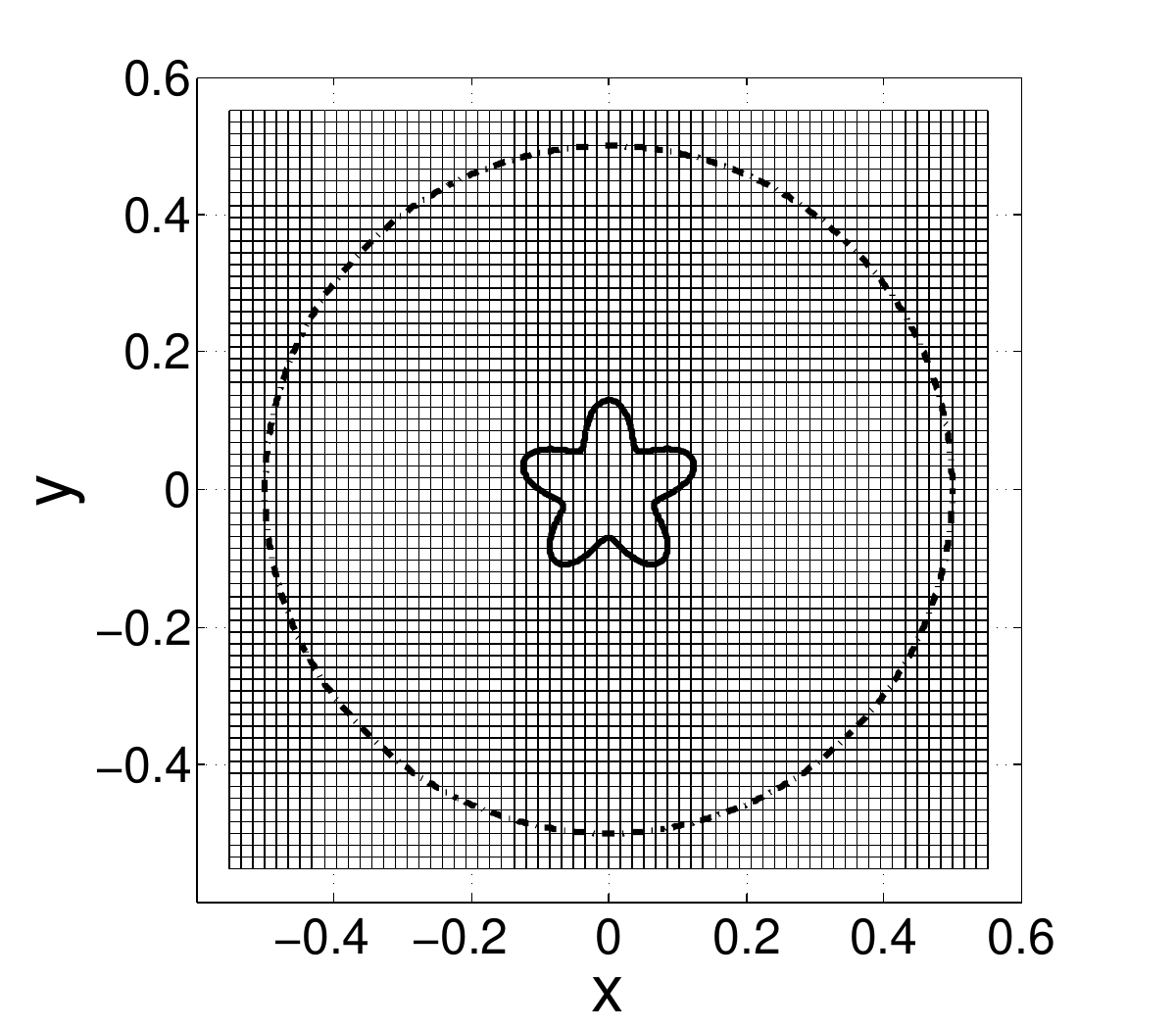}
  \includegraphics[width=2.6in]
  {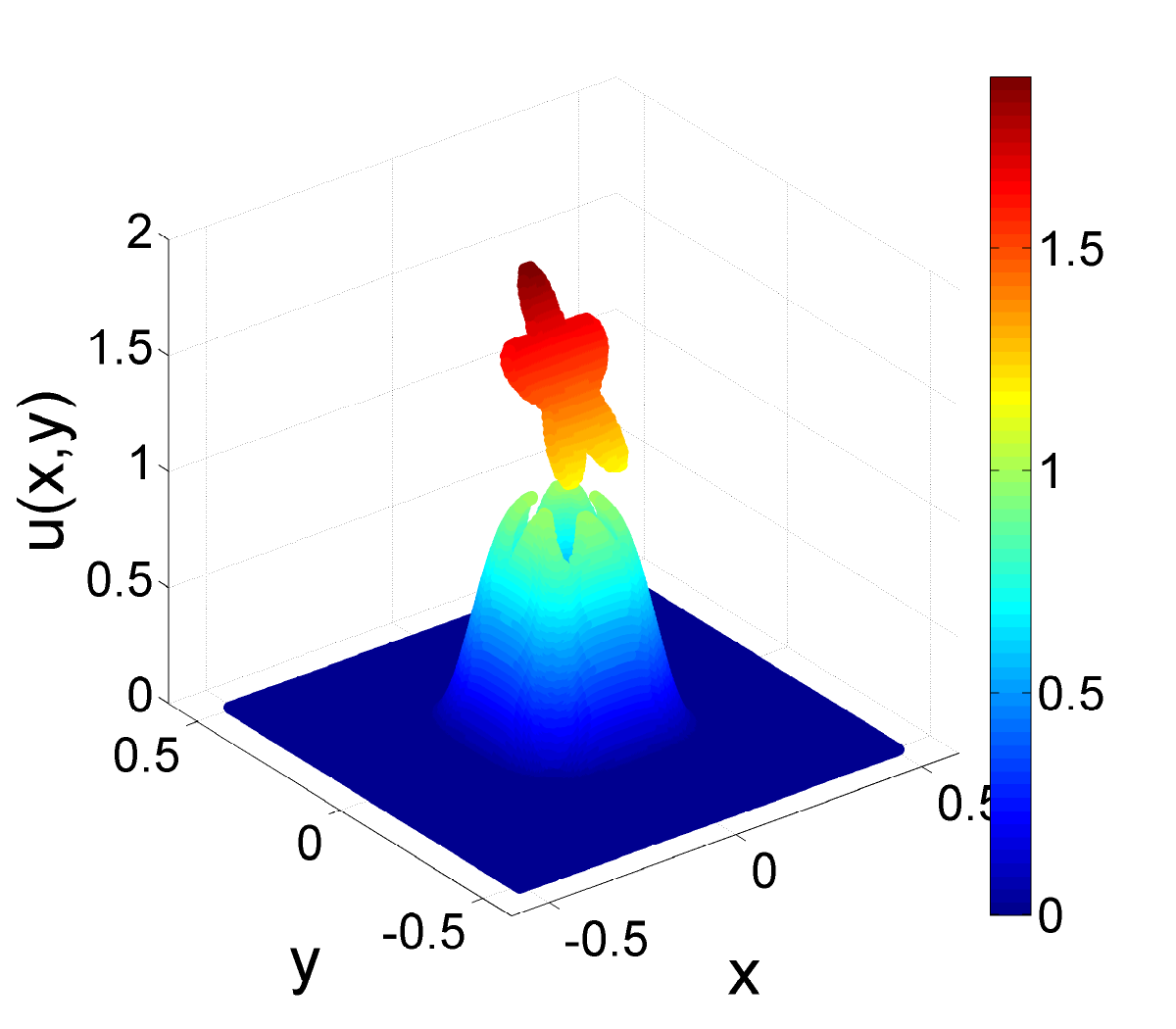}
 \end{center}
 \caption{Left: computational box $\mathcal{B}\/$ used in
   example 2. The solid black line represents the interface
   $\Gamma\/$, and the dashed line is the circle that
   defines the support of the source term. Right: solution
   obtained with the algorithm proposed in this paper.}
 \label{fig:ex2:grid}
\end{figure}

Note that the source term decays rapidly, such that we can
consider $f^+(r) \approx 0\/$ for $r > 0.4\/$. Hence, we set
$\Omega_f = \{r | r \le 0.4\}\/$ and solve the problem in a
square region slightly larger than $\Omega_f\/$.

In addition, case (ii) is an example of the poorly
conditioned problems discussed in
\S\ref{sub:problem:poorcond}. Thus, a direct application of
any solution algorithm can lead to poor results in this
case. For this reason we apply the prescription
in \S\ref{sub:solution:poorcond} and impose the constraint:
\begin{equation}\label{eq:constraint2}
 \int_{\Gamma} \rho_{\Gamma}\,d\/S = -2.4846\/.
\end{equation}
As shown in \S\ref{sub:solution:poorcond}, this constraint
can be computed with the original problem's data.
\revc{The right hand side in \eqref{eq:constraint2} is
computed using Gaussian quadrature, and the constraint is
implemented with the four digits of accuracy shown above.}
Figure~\ref{fig:ex2:grid} shows the solution to case (ii)
computed with \eqref{eq:constraint2}. This solution is
visually  indistinguishable from the solution obtained in
case (i).

\begin{figure}[htb!]
 \begin{center}
  \includegraphics[width=2.6in]
  {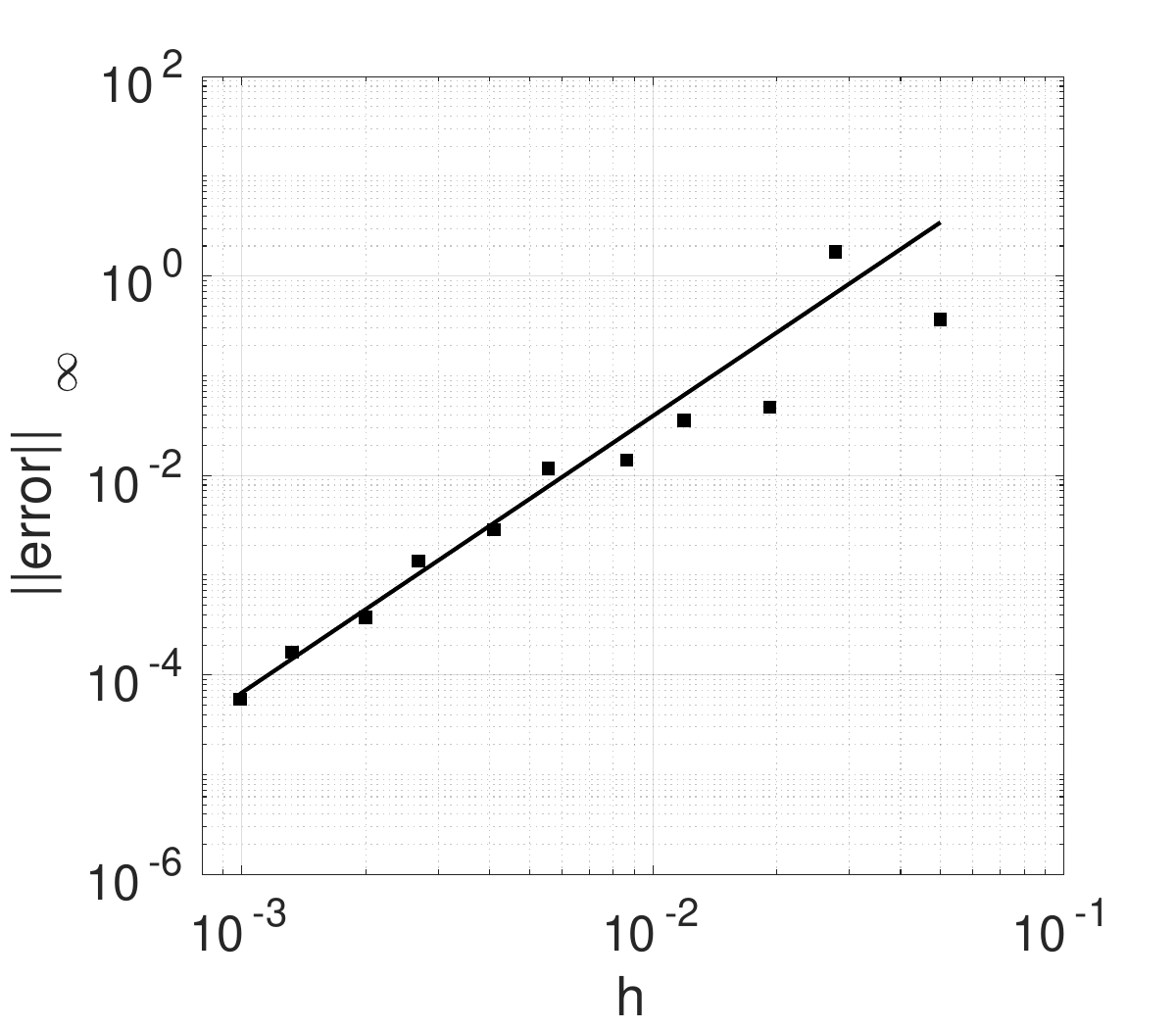}
  \includegraphics[width=2.6in]
  {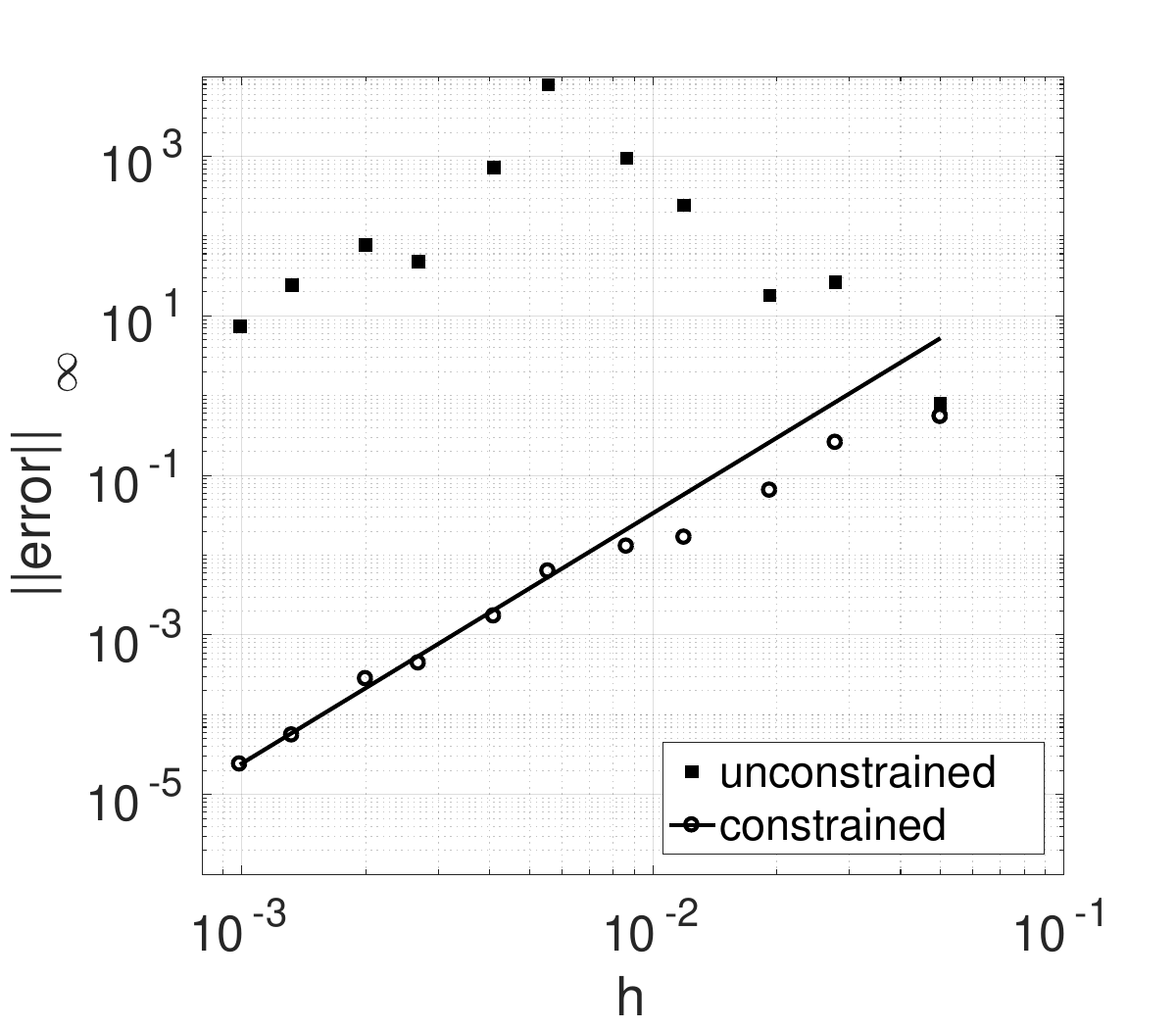}\\
  \hspace*{0.1in}
  \mbox{(a) case (i): $\beta^- = 1\/$, $\beta^+ = 10^6\/$.}
  \hfill
  \mbox{(b) case (ii): $\beta^- = 10^6\/$, $\beta^+ = 1\/$.}
  \hspace*{0.1in}  
 \end{center}
 \caption{Error convergence: $L_{\infty}\/$ norm of
    the error for the Poisson problems with discontinuous
    coefficients in example 2. The straight lines were
    obtained by least squares fits and their slopes
    correspond to the measured convergence rates -- see
    table~\ref{tab:convergence}.}
 \label{fig:ex2:convergence}
\end{figure}

Figure~\ref{fig:ex2:convergence} shows the convergence of
the error in the $L_{\infty}\/$ norm in both cases (i) and
(ii).
In case (ii) we plot the error obtained for the constrained
and unconstrained problems.
The measured convergence rates are listed in
table~\ref{tab:convergence} at the end of this section.
As expected, the measured rates are \trd\ order for both
cases: (i) and the constrained problem in (ii).
The solution to the unconstrained problem results in
$\mathcal{O}(1)\/$ errors, even though the residual produced
by GMRES is smaller than $10^{-10}\/$ in all cases.
This example shows how sensitive to errors the unconstrained
problem becomes when $\beta^-/\beta^+ \gg 1\/$.
In calculations with smaller beta-ratios the convergence of
the unconstrained problem can be ascertained to be \trd\
order.

\begin{figure}[htb!]
 \begin{center}
  \includegraphics[width=2.6in]
  {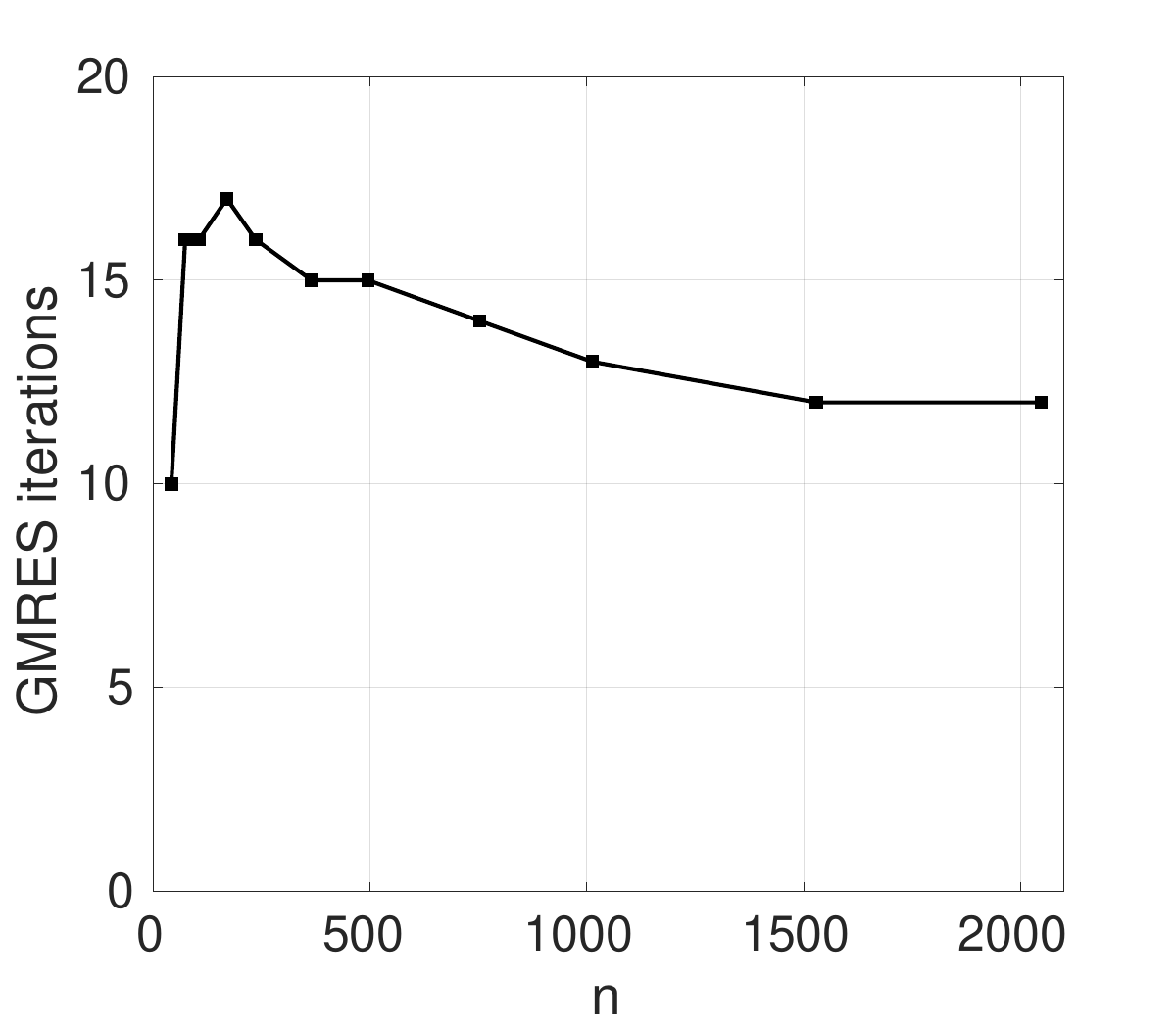}
  \includegraphics[width=2.6in]
  {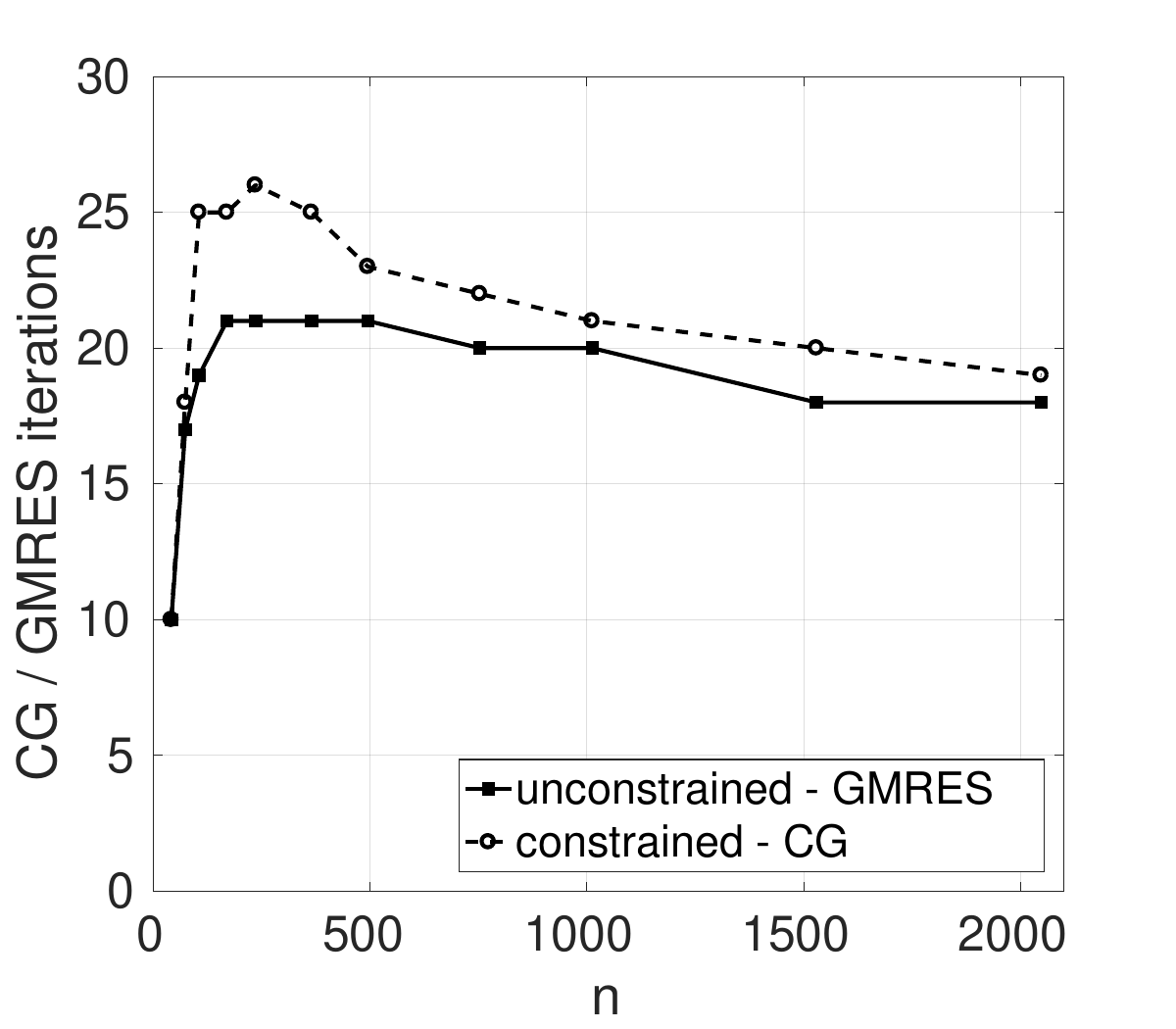}\\
  \hspace*{0.1in}
  \mbox{(a) case (i): $\beta^- = 1\/$, $\beta^+ = 10^6\/$.}
  \hfill
  \mbox{(b) case (ii): $\beta^- = 10^6\/$, $\beta^+ = 1\/$.}
  \hspace*{0.1in}  
 \end{center}
 \caption{\revc{Convergence of Krylov solvers: number of
    iterations needed to compute the potential distributions
    at the interfaces $\Gamma\/$ and $\partial \Omega_f\/$.
    Here $n$ indicates the total number of markers used to
    discretize both interfaces.}}
 \label{fig:ex2:GMRES}
\end{figure}

\revc{Finally, figure~\ref{fig:ex2:GMRES} shows the number of
iterations the Krylov solver (either GMRES or CG) requires
to converge to a residual tolerance of $10^{-10}$.
The Krylov solver is used to compute the potential
distribution of the boundary integral formulation.
The unconstrained problem results in a non-symmetric linear
system that is solved using GMRES.
On the other hand, the constrained problem results in an
overdetermined linear system, and the normal equation is
solved using CG.
Figure~\ref{fig:ex2:GMRES} shows that the Krylov solvers
converge with a relatively small number of iterations, and
that the number of iterations that does not depend on the
size of the problem.
Note that no pre-conditioners are used with GMRES and CG.}

\subsection{Example 3. Interior Dirichlet and Neumann
            Poisson problems with a discontinuity interface}
\label{sub:results:interior_interface}
\begin{figure}[htb!]
 \begin{center}
  \includegraphics[width=2.6in]
  {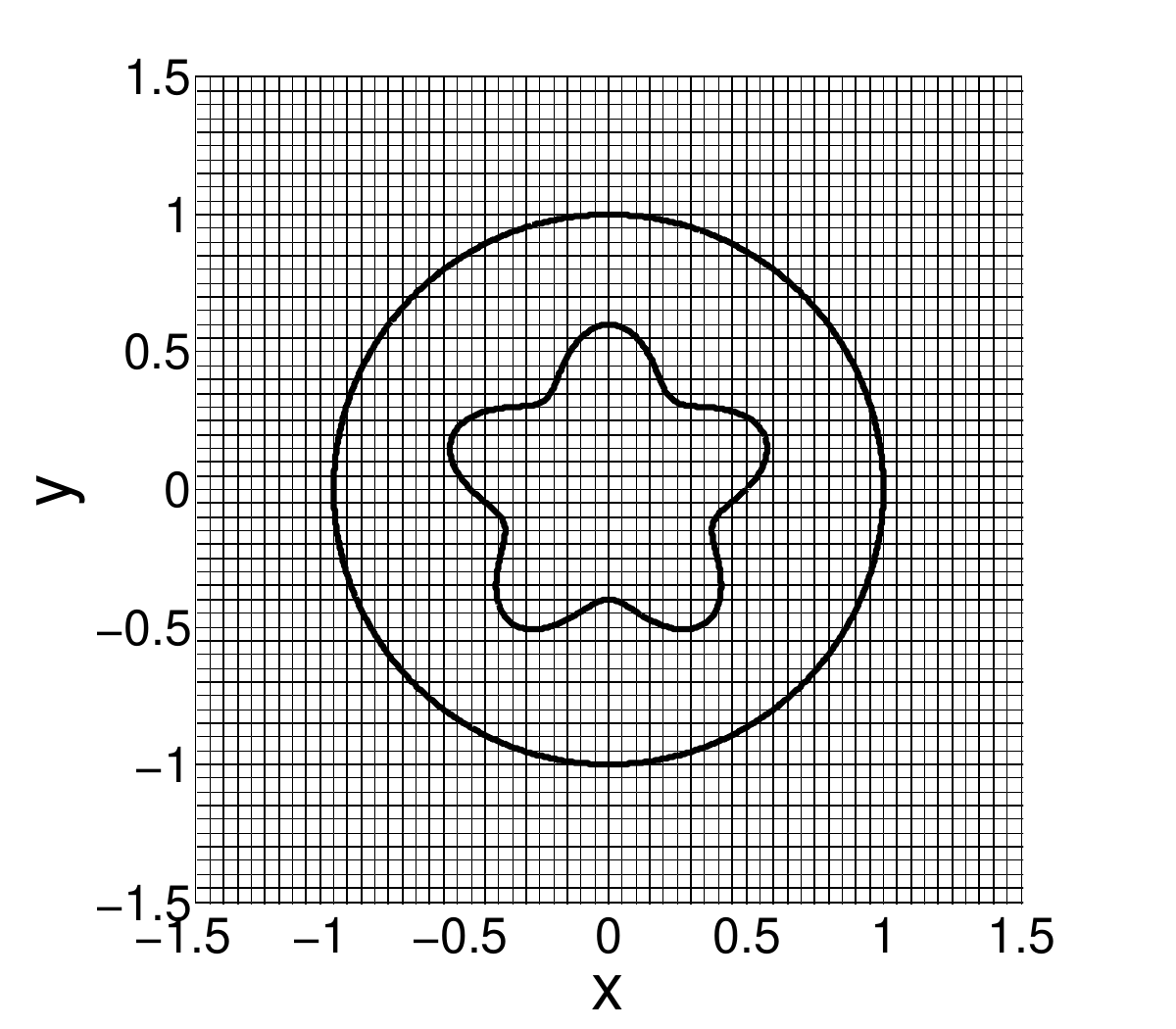}
  \includegraphics[width=2.6in]
  {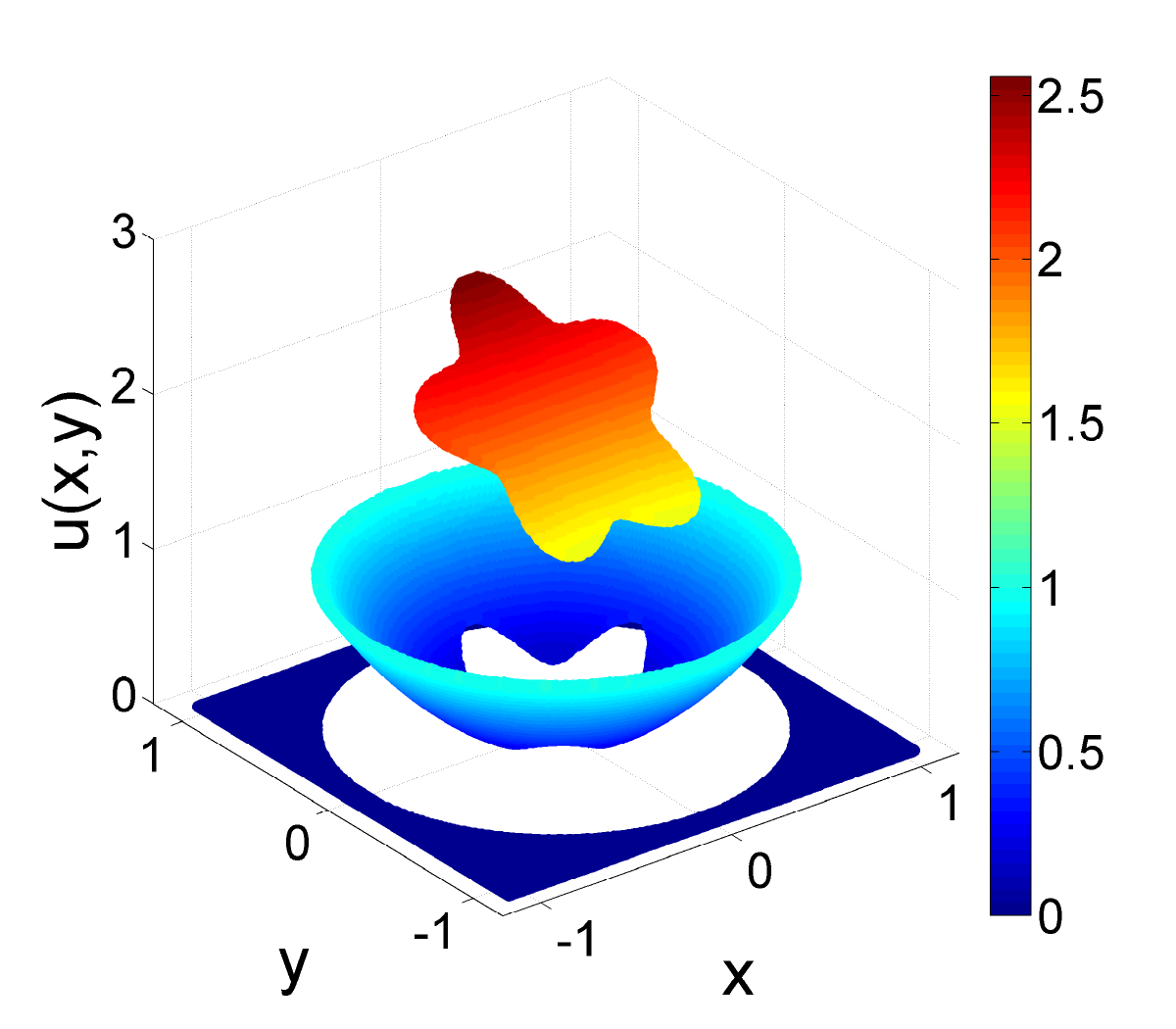}\\
 \end{center}
 \caption{Left: solution domain $\Omega\/$ for example 3,
    embedded in a Cartesian Grid. The boundary
    $\partial \Omega\/$, and the discontinuity interface
    $\Gamma\/$, are the thick solid lines. Right: solution
    obtained with the algorithm proposed in this paper.}
 \label{fig:ex3:grid}
\end{figure}

In this example, the solution domain is the unit disk, while
the discontinuity interface is defined by \eqref{eq:star}
with parameters $\delta=0.1\/$ and $r_0=0.5\/$.
Figure~\ref{fig:ex3:grid} shows the solution domain and the
interface immersed in a regular Cartesian grid. In this
domain, we solve the interior Dirichlet and Neumann Poisson
problems associated with the exact solution
\begin{subequations}\label{eq:ex3}
 \begin{align}
  u^+(x,\,y) &= x^2 + y^2\/,\\ \rule{0mm}{1.2em}
  u^-(x,\,y) &= \cos\/(x)\,\sin\/(y) + 2\/.
 \end{align}
\end{subequations}
Furthermore, as in \S\ref{sub:results:open}, we consider two
cases in which the coefficients $\beta^{\pm}\/$ have large
jumps across the discontinuity interface:
\begin{enumerate}[(i)]
 \item $\beta^- = 1\/$, $\beta^+ = 10^6\/$,
 \item $\beta^- = 10^6\/$, $\beta^+ = 1$.
\end{enumerate}
Case (ii) is another instance of the poorly conditioned
problem discussed in \S\ref{sub:problem:poorcond}.
Hence, in this example we apply the prescription
in~\S\ref{sub:solution:poorcond} and impose the constraint:
\begin{equation}\label{eq:constraint3}
 \rule{0mm}{1.7em}
 \int_{\Gamma} \rho_{\Gamma}\,d\/S = 3.2044\/.
\end{equation}
\revc{The right hand side in \eqref{eq:constraint3} is
computed using Gaussian quadrature, and the constraint is
implemented with the four digits of accuracy shown above.}

In addition, the solutions to the Neumann problems are only
defined up to an arbitrary constant. GMRES will
automatically ``pick'' a constant, which may not correspond
to \eqref{eq:ex3}. Hence, we shift the Neumann solutions to
obtain $u^-(0,0) = 2\/$. After this shift, the Dirichlet and
Neumann solutions to case (i), and case (ii) with
\eqref{eq:constraint3}, become visually indistinguishable.
Figure~\ref{fig:ex3:grid} shows the solution to the
constrained Neumann problem in case (ii). 

\begin{figure}[htb!]
 \begin{center}
  \includegraphics[width=2.6in]
  {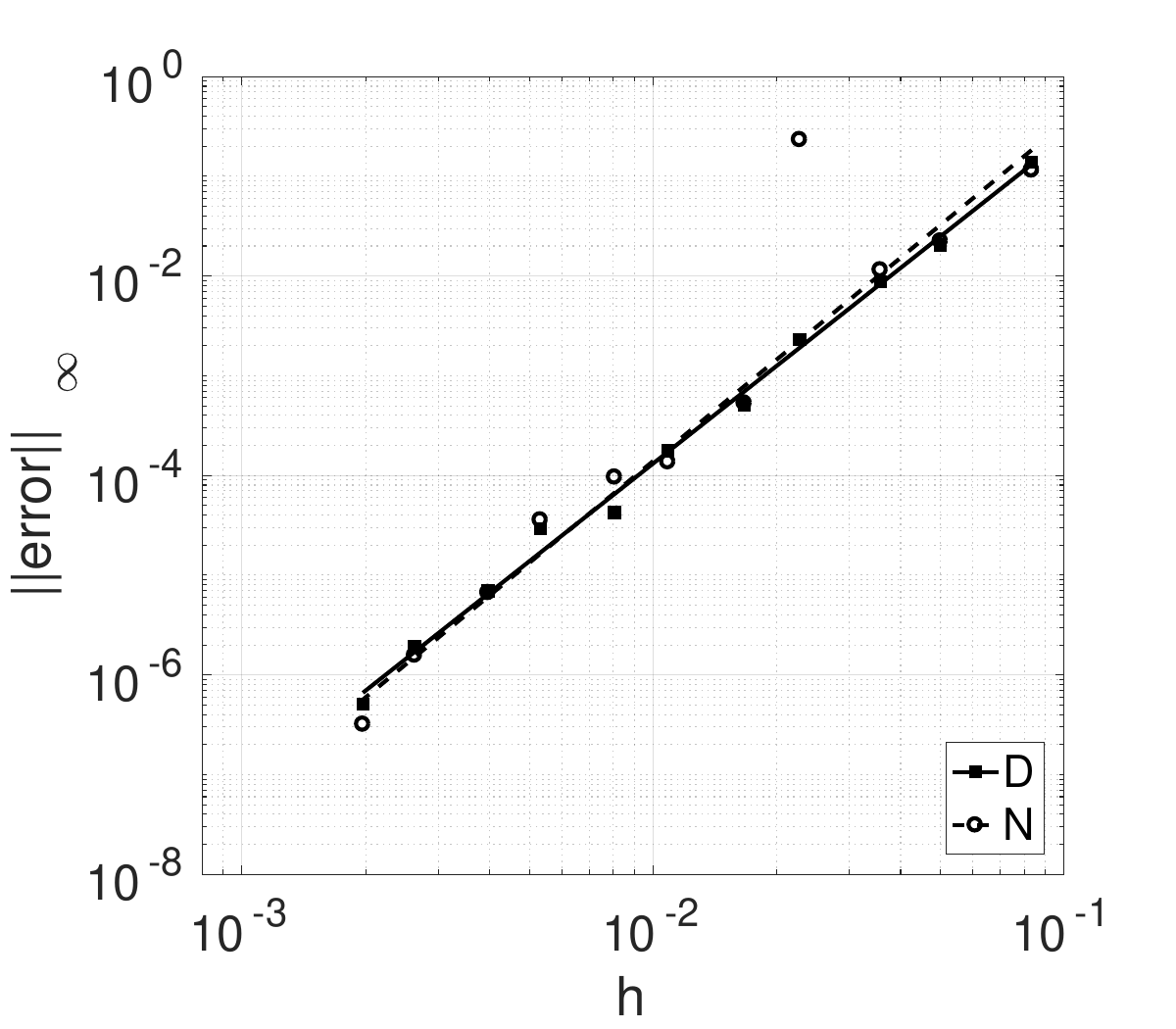}
  \includegraphics[width=2.6in]
  {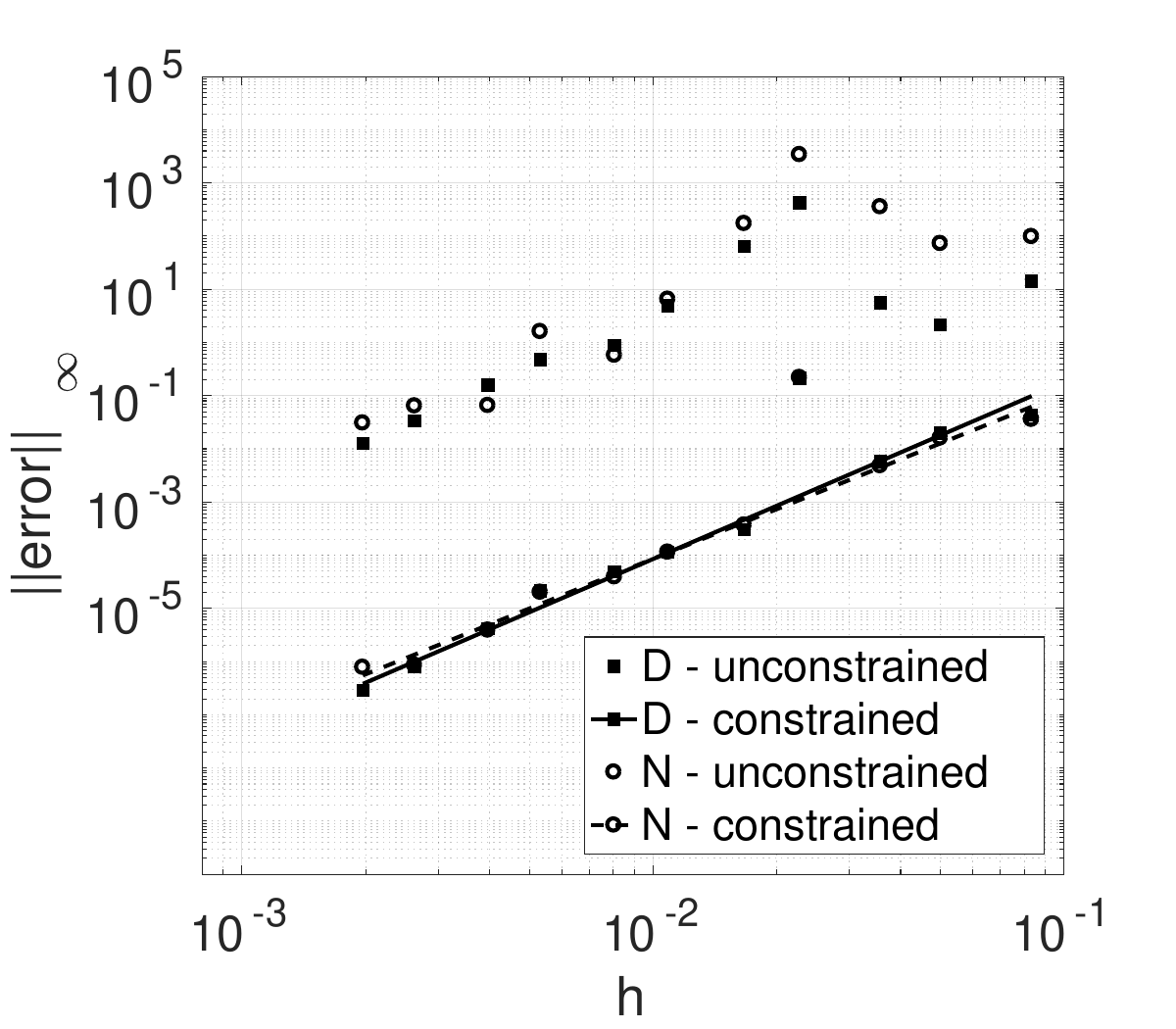}\\
  \hspace*{0.1in}
  \mbox{(a) case (i): $\beta^- = 1\/$, $\beta^+ = 10^6\/$.}
  \hfill
  \mbox{(b) case (ii): $\beta^- = 10^6\/$, $\beta^+ = 1\/$.}
  \hspace*{0.1in}  
 \end{center}
 \caption{\rrrc{Error convergence: $L_{\infty}\/$ norm of
    the error for the Poisson problems with discontinuous
    coefficients in example 3.} The straight lines were
    obtained by least squares fits and their slopes
    correspond to the  measured convergence rates -- see
    table~\ref{tab:convergence}.}
 \label{fig:ex3:convergence}
\end{figure}

The convergence of error, measured in the $L_{\infty}\/$
norm, is displayed in figure~\ref{fig:ex3:convergence}. In
case (ii) we plot the error obtained for the constrained
and unconstrained problems. The measured convergence rates
are listed in table~\ref{tab:convergence} at the end of this
section. As expected, we observe \trd\ order convergence for
case (i) and the constrained problem of case (ii).
Furthermore, the errors obtained for the unconstrained
problem are significantly larger than the errors obtained
for the constrained problem in case (ii).

\begin{figure}[htb!]
 \begin{center}
  \includegraphics[width=2.6in]
  {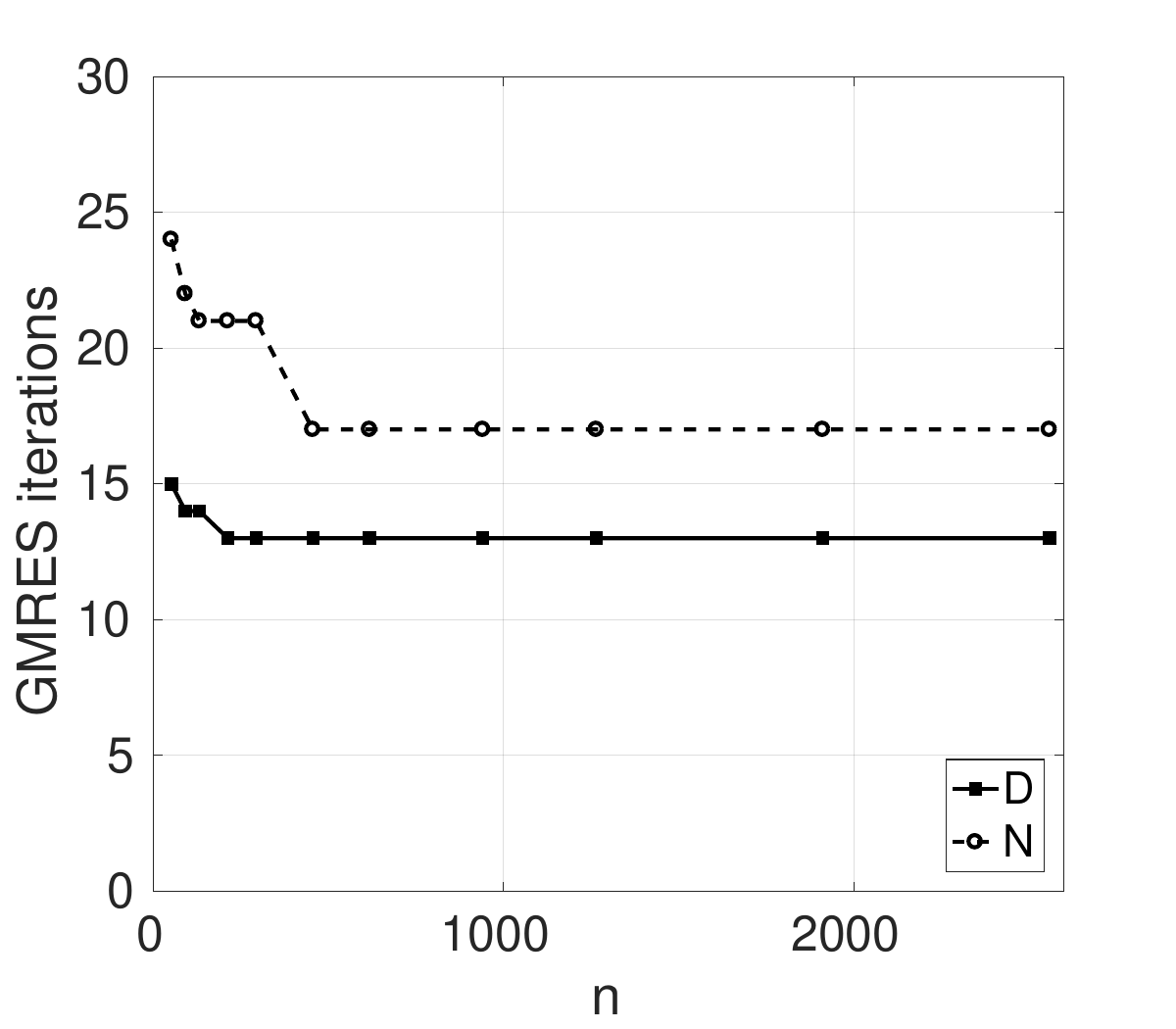}
  \includegraphics[width=2.6in]
  {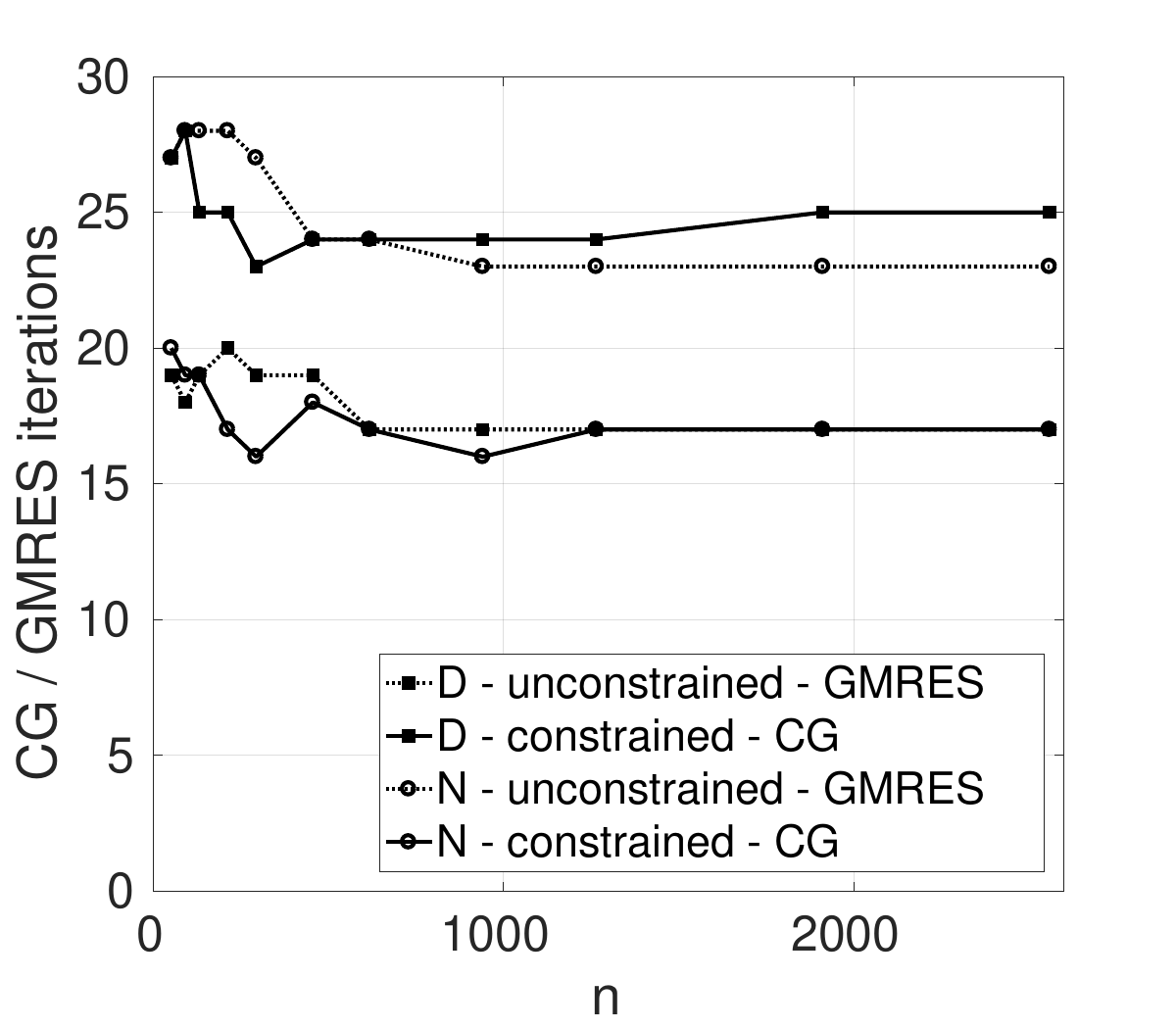}\\
  \hspace*{0.1in}
  \mbox{(a) case (i): $\beta^- = 1\/$, $\beta^+ = 10^6\/$.}
  \hfill
  \mbox{(b) case (ii): $\beta^- = 10^6\/$, $\beta^+ = 1\/$.}
  \hspace*{0.1in}  
 \end{center}
 \caption{\revc{Convergence of Krylov solvers: number of
    iterations needed to compute the potential distributions
    at the interfaces $\Gamma\/$ and the boundary
    $\partial \Omega\/$.
    Here $n$ indicates the total number of markers used to
    discretize both the interface and the boundary.}}
 \label{fig:ex3:GMRES}
\end{figure}

\revc{Finally, figure~\ref{fig:ex3:GMRES} shows the number of
iterations the Krylov solver (either GMRES or CG) requires
to converge to a residual tolerance of $10^{-10}$.
As in previous examples, no preconditioners were used.
Once again we observe that the Krylov solvers converge with
a relatively small number of iterations, and that the number
of iterations that does not depend on the size of the
problem.}

\begin{table}
 \caption{\revc{Measured convergence rate of the error in the
    $L_\infty$ norm for examples 1 through 3 --- see
    figures~\ref{fig:ex1:convergence},
    \ref{fig:ex2:convergence}, and
    \ref{fig:ex3:convergence}.
    Error limits correspond to one standard deviation
    of the least squares error.}
    Only solutions with error smaller than 0.01 were
    considered, since the measured convergence rates serve
    as estimates of the asymptotic order of accuracy of the
    algorithm.}
 \label{tab:convergence}          
 \begin{center}
  \begin{tabular}{l|c|c}
   \hline\hline
   \multirow{3}{*}{\textbf{Problem}}
   & \textbf{Measured} & \textbf{Theoretical}\\
   & \textbf{convergence} & \textbf{convergence}\\
   & \textbf{rate} & \textbf{rate}\\
   \hline
   Ex.\ 1 -- Dirichlet & $3.8 \pm 0.1$ & 4.0\\
   Ex.\ 1 -- Neumann & $3.2 \pm 0.1$ & 3.0\\
   Ex.\ 2 -- case (i) & $2.8 \pm 0.2$ & 3.0\\
   Ex.\ 2 -- case (ii) constrained & $3.1 \pm 0.1$ & 3.0\\
   Ex.\ 3 -- case (i) -- Dirichlet & $3.3 \pm 0.1$ & 3.0\\
   Ex.\ 3 -- case (i) -- Neumann & $3.4 \pm 0.3$ & 3.0\\
   Ex.\ 3 -- case (ii) constrained -- Dirichlet & $3.3 \pm 0.1$ & 3.0\\
   Ex.\ 3 -- case (ii) constrained -- Neumann & $3.1 \pm 0.1$ & 3.0 \\
   \hline
  \end{tabular}
 \end{center}
\end{table}


%% file: 6conclusion.tex
\section{Conclusion}\label{sec:conclusion}
In this paper we presented an algorithm to solve Poisson
problems in arbitrarily shaped domains immersed in regular
Cartesian grids. The algorithm can be applied to a wide
range of Poisson problems, including cases with prescribed
internal jumps on both the solution and the weighted normal
derivatives (with prescribed weights on each side). In
particular, we showed that the algorithm produces good
results even when the ratio between the weights
is very large --- as is the case in many multi-phase flow
applications.

Mayo and collaborators~\cite{mayo:84,
mayo:jcp:1992, mckenney:95} proposed a similar method for the
Laplace and Poisson equations in an immersed setting.
However, they considered only the case of immersed
boundaries, and not interfaces of discontinuity.
Furthermore, extensions of their method to general order of
accuracy do not seem to be straightforward. In contrast,
because we use the Correction Function Method (CFM), our
method (at least in principle) can be readily extended to
any desired order of accuracy.

The algorithm presented here relies on efficient and
accurate numerical methods. The algorithm is comprised of
three sub-problems: a boundary integral equation, and two
``constant coefficients'' Poisson problems defined in
rectangular domains\,\footnote{The
 meaning that ``constant coefficients'' has in this context
 is explained in remark~\ref{rmk:problem:ccoeff}.}.
The boundary integral equation is solved using well
established boundary integral methods (BIM), whereas the
Poisson equations are solved with finite differences and the
CFM.

The accuracy of the algorithm was illustrated with several
2-D examples. The version of the algorithm implemented
achieved \trd\ or \fth\ order of accuracy, depending on the
characteristics of the problem. However, the
BIM and CFM may be implemented to any order of accuracy.
Hence, there is no inherent limit to the accuracy that can
be achieved with the algorithm proposed here.

Finally, the algorithm can be extended to 3-D. We have not
yet implemented the algorithm in 3-D because the evaluation
of the integrals in equation~\eqref{eq:w2:bc} becomes an
efficiency bottleneck. However, recent work~\cite{ying:13}
on the Kernel-Free Boundary Integral
Method (KFBIM) provides a natural path for addressing this
issue. Currently we are investigating the joint use of the
CFM and the KFBIM for the purpose of obtaining an algorithm
that is both high order (as the algorithm in this paper),
and also efficient in 3-D.

%% file: acknowledgements.tex
\section*{Acknowledgements}
The work by A.\ N.\ Marques was partially supported by
CAPES (Coordena\c{c}\~ao de Aperfei\c{c}oamento de Pessoal
de N\'ivel Superior, Brazil) and the Fulbright Commission
via the joint grant BEX 2784/06-8. The work by J.-C.\ Nave
was partially supported by the NSERC Discovery and Discovery
Accelerator programs. Finally, the work by R.\ R.\ Rosales
was partially supported by NSF grants DMS-1318942 and
DMS-1614043.